\documentclass{article}
\usepackage{geometry}
\usepackage{amsmath}
\usepackage{amsfonts}
\usepackage{amstext}
\usepackage{amssymb}
\usepackage{amsthm}
\usepackage{eufrak}
\usepackage{color}
\usepackage{graphicx}

\newtheorem{lemma}{Lemma}

\newtheorem{theorem}{Theorem}
\newtheorem{corollary}{Corollary}
\theoremstyle{definition}
\newtheorem{definition}{Definition}

\theoremstyle{remark}
\newtheorem{remark}{Remark}

\begin{document}

\title{Small-time asymptotics for Gaussian self-similar stochastic
volatility models}
\author{Archil Gulisashvili,\thanks{%
Department of Mathematics, Ohio University, Athens, Ohio 45701. Email: 
\texttt{gulisash@ohio.edu}} \quad Frederi Viens,\thanks{%
Department of Statistics, Purdue University, West Lafayette, IN 47907.
Email: \texttt{viens@purdue.edu}} \quad and Xin Zhang,\thanks{%
Department of Mathematics, Purdue University, West Lafayette, IN 47907.
Email: \texttt{zhang407@math.purdue.edu}}}
\date{}
\maketitle

\begin{abstract}
We consider the class of self-similar Gaussian stochastic volatility models,
and compute the small-time (near-maturity) asymptotics for the corresponding
asset price density, the call and put pricing functions, and the implied
volatilities. Unlike the well-known model-free behavior for extreme-strike
asymptotics, small-time behaviors of the above depend heavily on the model,
and require a control of the asset price density which is uniform with
respect to the asset price variable, in order to translate into results for
call prices and implied volatilities. Away from the money, we express the
asymptotics explicitly using the volatility process' self-similarity
parameter $H$, its first Karhunen-Lo\`{e}ve eigenvalue at time 1, and the
latter's multiplicity. Several model-free estimators for $H$ result. At the
money, a separate study is required: the asymptotics for small time depend
instead on the integrated variance's moments of orders $\frac{1}{2}$ and $%
\frac{3}{2}$, and the estimator for $H$ sees an affine adjustment, while
remaining model-free.
\end{abstract}

\vspace{0.2in%
}

\vspace{0.2in}

\bigskip

\bigskip

\section{Introduction}

In this paper, we present a study of the small-time (near-maturity)
asymptotics for the asset price density $S$, the call and put prices, and
the implied volatilities, for the class of continuous-time
Black-Scholes-Merton-type models with Brownian noise and independent
Gaussian self-similar volatility. The techniques borrow from a framework
established in our prior work \cite{GVZ} for general Gaussian volatility
models; they use a tailored application of Laplace's method requiring a
delicate analysis of uniformity with respect to strike prices $K$ away from
the money ($K\neq s_{0}$), and apply a general result from \cite{GL} to
translate asymptotics from call prices to implied volaltilities. Model-free
estimators of the self-similarity parameter $H$ result. Away from the money,
all asymptotic constants and powers are expressed explicitly in terms of $H$
and of the coefficients in the Karhunen-Lo\`{e}ve expansion of the
volatility. At the money ($K=s_{0}$), a separate study is required. This
introduction contains extensive details of general context of the small-time
asymptotic problems mentioned above, our motivations, and a precise summary
of all our results.

\subsection{General background}

It has been known for decades that the Bachelier-Black-Merton-Scholes
framework, while extraordinarily fertile for explaining various basic
features of financial markets and for helping define fundamental notions,
including volatility as the relative scale of noise intensity, suffers from
certain deficiencies, particularly the fact that volatility is not constant
empirically. When coupled with the fact that non-random volatility, which
implies normally distributed log returns, has difficulties in explaining
certain extreme events because of excessively light tails, one quickly
arrives at the vast class of stochastic volatility models, i.e. those
continuous-time models where the relative noise intensity of returns is
itself a stochastic process which is at least partially driven by exogenous
noise. A large number of articles and monographs on stochastic volatility
(SV) can be consulted for empirical and economic justification of these
models; we cite the classical text \cite{FPS}. Of particular interest is SV
models' ability to reproduce some desirable market features of option
prices, such as \textquotedblleft smiles\textquotedblright\ and other
non-flat shapes of the implied volatility (IV), i.e. the volatility which
would be required of a constant-volatility model to explain a given call
option price.

One of the first mathematical treatments explaining empirically observed IV
shapes was by Renault and Touzi in \cite{RT}. Recent studies have looked in
detail at the question of IV asymptotics, that is to say the behavior of IV
as important parameters such as strike price $K$ and maturity $T$ tend to
extreme values. Of note is the groundbreaking paper \cite{Lee} of Lee, in
which the large-strike (the small-strike) behavior of IV is described in
terms of the largest (the smallest) non-exploding moment 
of the stock price. Gaussian volatility models belong to the class of models
with moment explosions. For more details and other references on IV shapes
and extreme-strike asymptotics of IV, we refer to the introduction section
in our prior work \cite{GVZ}, where we examine the class of uncorrelated
Gaussian volatility models in its broadest possible sense.

\subsection{Specific motivations and modeling choices}

Small-time asymptotic behavior of densities, option pricing functions, and
implied volatilities has been a popular topic of study. There are various
model-independent results (see, e.g., \cite{CC,GL,HL,RR}), explaining how
the asymptotics of the IV depend on those of option pricing functions. There
are also papers discussing small-time asymptotics of the functions mentioned
above in the case of stochastic volatility or local-stochastic volatility
models (see \cite{AFLZ,BBF,FFK,FJ2,FJ3,GHLOW,HL,P}), and for special models
(see \cite{ALV,FH,FF,MS,MKN,MT} (models with jumps), \cite{FFF,FJ1,FJL,FJM}
(Heston model), \cite{D1,D2} (Stein-Stein model), \cite{HKLW,HLW,GHJ,P}
(SABR model)).

The present paper follows up on our prior study in \cite{GVZ} by attempting
to elucidate the small-time behavior of IV for a subclass of Gaussian
volatility models, consisting of models with self-similar volatility
processes. It turns out that establishing small-time asymptotics in a
general Gaussian context is significantly more demanding than determining
large-strike behavior. This can be understood as a manifestation of the fact
that there is no model-free analogue of Lee's moment formulas in the small
or large time regimes. In this paper, we illustrate the challenge by
specializing to the case of self-similar volatilities; we will see that the
type of small-time behavior for both call price and IV is quite sensitive to
the self-similarity parameter $H$. This is good news if one is to leverage
these results to help determine $H$, as we will see.

Indeed, our study also allows us to investigate the question of long-memory
SV calibration, since long-range dependence and self-similarity are proxies
for each other in many known models, via their common Hurst parameter $H$.
Based on a Gaussian long-memory model for log-volatility pioneered by Comte
and Renault in \cite{CR}, the work in \cite{CV} used an ad-hoc calibration
method based on option prices to determine $H$ so as to best explain market
prices. Fractional volatility models also appear in \cite%
{BFG,CCR,JG1,JG2,GJRrough,FZ,Fu1,Fu,GJR,MOR,R}. In the current paper, we
show that calibration of $H$ near maturity can be given a stronger
mathematical foundation under self-similarity assumptions for the volatility
process. The parameter $H$ can also be a proxy for local regularity
measurements, in the sense of their paths' H\"{o}lder continuity parameter.
Some recent papers and presentations, yet unpublished at the time of writing
this article, appear to show that volatility is rough, in the sense that the
log-volatility process is fractional and it is not H\"{o}lder continuous for 
$1/2-\varepsilon <H<1/2$, where $\varepsilon $ is a positive number 
(see \cite{JG1, JG2, GJRrough}). On the other hand, \cite{CV} and many
studies before it (see references therein) indicate that $H>1/2$ in terms of
memory length. This is a demonstration that the use of $H$ to measure
self-similarity \emph{and} long memory \emph{and} path regularity/roughness,
such as in the case of fractional Brownian motion (fBm), might be a
misspecification in volatility modeling. The authors of \cite{GJRrough}
indicate that classical long-memory tests detect this property in their
Gaussian rough volatility model, which is a geometric fBm or a geometric OU
process with shorter memory ($H<1/2$). The studies in \cite{CV} show on the
other hand that no consistent memory estimation results in practice from any
classical method when used on the non-self-similar stationary long-memory
model of \cite{CR}. Our current work could help in elucidating the
differences between these points of view; we do not comment on them further
herein. An interesting discussion of long memory vs short memory problem can
be found in Section 1.2 of \cite{GJRrough}. In any case, the numerics which
we include in this paper and will discuss at the end of this introduction
show that our model class allows for a very sharp calibration tool.

Before providing a summary of our results, we discuss some classical
Gaussian self-similar models. General details about this class are given in
Section \ref{S:dva}. These are the Gaussian processes $X$ on $[0,T]$ such
that for some $H\in (0,1)$ and for any $a>0$, the two processes $t\mapsto
X_{at}$ and $t\mapsto a^{H}X_{t}$ have the same distribution (law). The best
known among them is the fractional Brownian motion (fBm) $B^{H}$, the
centered Gaussian process whose law is defined by $B^{H}\left( 0\right) =0$
and $\mathbf{E}\left[ \left( B_{t}^{H}-B_{s}^{H}\right) ^{2}\right]
=\left\vert t-s\right\vert ^{2H}$. It is the only (continuous) self-similar
centered Gaussian process with stationary increments. Many texts can be
consulted on $B^{H}$, including, e.g., \cite{No,Nu,R}. Among the many other
centered Gaussian self-similar models, which are all necessarily
non-stationary, the easiest to construct is the Riemann-Liouville fBm,
defined as $B_{t}^{H,RL}=\int_{0}^{t}\left( t-s\right) ^{H-1/2}dW\left(
s\right) $ where $W$ is a standard Wiener process (see for instance \cite%
{Lim}). This process, which is $H$-self-similar, has properties close to
those of fBm, and can be more amenable to calculations. The so-called
Bifractional Brownian motion depends on two similarity parameters $H$ and $K$%
, has a more complex representation, as the sum of an fBm with parameter $HK$%
, and a process with $C^{\infty }$ paths which is not adapted to a Brownian
filtration: see \cite{HV}, see also \cite{BES} and the references therein.
This process, which is $HK $-self-similar, can model the effect of smoothly
acquired exogenous information, and is an extension of the so-called
sub-fractional Brownian motion (see \cite{BGT}). Self-similar Gaussian
processes can also be obtained as the solutions of stochastic partial
differential equations: a class which includes solutions to fractional
colored stochastic heat equations is studied in \cite{TTV}, which has the
interesting property that its discrete quadratic variation has fluctuations
which become non-Gaussian at a threshold of self-similarity which is lower
than for fBm, and can be adjusted to be as low as desired. This can be
helpful to model volatilities whose local behavior has heavier-tailed
fluctuations than what standard fBm can allow, regardless of the
volatility's self-similarity. It also allows the modeler to choose
regularity and self-similarity properties independently of each other, which
offers more flexibility than the models considered in \cite{CR, CV, GJRrough}%
. More examples of Gaussian self-similar process can be found in \cite%
{BGT,DM}. Interestingly, many of the Gaussian self-similar models share the
same path regularity properties as fBm, because it can be shown that there
are positive finite constants $c,C$ for which $c\left\vert t-s\right\vert
^{2H}<\mathbf{E}\left[ \left\vert X_{t}-X_{s}\right\vert ^{2}\right]
<C\left\vert t-s\right\vert ^{2H}$, where the symbol $H$ stands for the
self-similarity parameter of the model under consideration.

Finally, it bears noting that self-similarity implies that $X_{0}=0$ and
that $Var\left[ X_{t}\right] $ is proportional to $t^{2H}$. This is a strong
assumption on $X$. An uncertainty level on volatility which increases with
time is a reasonable conservative forecasting assumption. That the volatilty
starts at $0$ is more restrictive, since, in our IV context, it corresponds
to saying that the underlying risky asset's movements tends towards
certainty near the derivative's maturity. Such a behavior is characteristic
of specific risky asset classes, such as fixed-income securities, e.g.
treasury bonds, and the dividend streams in preferred stocks; it is atypical
of common stocks. To soften the assumption that $X_{0}=0$, one can add a
constant mean to each centered self-similar $X$. We have investigated this
possibility; it appears that this will require additional non-trivial tools
not contained herein. Given the length of the current article, we have opted
to leave this improvement for another work. One may, however, include a
non-zero mean for each $X_{t}$ which is proportional to $t^{H}$; this is the
framework used herein throughout.

\subsection{Summary of main results, proof techniques, and numerics}

In \cite{GVZ}, we studied Gaussian stochastic volatility models. The asset
price process $S$ in such a model satisfies the following linear stochastic
differential equation: 
\begin{equation}
dS_{t}=rS_{t}dt+|X_{t}|S_{t}dW_{t},  \label{E:svm}
\end{equation}%
where $X$ is a continuous adapted Gaussian process on a filtered complete
probability space $(\Omega ,\mathcal{F},\{\mathcal{F}_{t}\},\mathbb{P})$, $W$
is a standard Brownian motion on $(\Omega ,\mathcal{F},\mathbb{P})$ with
respect to the filtration $\{\mathcal{F}_{t}\}$, $S_{0}=s_{0}>0$ a.s., and $%
r\geq 0$ is the risk-free interest rate. We will assume throughout the paper
that the processes $X$ and $W$ are independent. In the model in (\ref{E:svm}%
), the volatility is described by the absolute value of a continuous
Gaussian process. An important special example of a Gaussian stochastic
volatility model is the Stein-Stein model introduced in \cite{SS}, where $X$
in (\ref{E:svm}) is an Ornstein-Uhlenbeck process.

If $S_{0}=s_{0}$, the call option on $S$ with maturity $T$ and strike price $%
K$ has price $C\left( T,K\right) $; this price equals a price $C_{BS}\left(
T,K;\sigma \right) $ in the Black-Scholes model with the volatility $\sigma $
depending on $T$ and $K$. That value of $\sigma $ is called the \emph{%
implied volatility} (IV) and is denoted by $I\left( T,K\right) $. In the
present paper, we concentrate on the behavior of $C$ and $I$ for small $T$
when $K$ is fixed; consequently, we typically drop the dependence of $C$ and 
$I$ on $K$.

Of particular importance is the density $p_{T}$ of the \emph{integrated
variance} $Y_{T}:=\int_{0}^{T}X_{t}^{2}dt$. The centered version of this $%
Y_{T}$ is a random variable in the second chaos of a Wiener space
independent of $W$. The covariance function of $X$ acts as a compact
self-adjoint linear operator on $L^{2}\left( [0,T]\right) $, with non-zero
eigenvalues $\left( \lambda _{n}:n=1,2,\ldots \right) $ arranged in
non-increasing order with repeats for multiplicities. This, and the
corresponding eigenfunctions, are the basis for the so-called Karhunen-Lo%
\`{e}ve (KL) decomposition of $X$ (see, e.g., \cite{A,Y}), and of a
corresponding one for $Y$. In any case, the asymptotic behavior of $p_{T}$
near $+\infty $, which was established in \cite{GVZ}, depends on specific KL
statistics, including the top eigenvalue $\lambda _{1}$, its multiplicity $%
n_{1}$, and the rescaled $L^{2}\left( [0,T]\right) $-orthogonal projection $%
\delta $ of $\lambda _{1}$'s eigenspace on the mean function of $X$ (see
Theorem \ref{T:gella} below). When applied to the case of $H$-self-similar $%
X $, via the simple scaling formula $p_{T}(y)=T^{-2H-1}p_{1}\left(
T^{-2H-1}y\right) $, the behavior of $p_{T}\left( \cdot \right) $ at $%
x\rightarrow +\infty $ translates into an expansion around $T\rightarrow
0^{+}$ of the density $\tilde{p}_{T}\left( x\right) $ of the rescaled
square-rooted version of $Y_{T}$ which is precise up to a factor $\left(
1+O_{x}\left( T^{H}\right) \right) $ for any fixed $x>0$: see asymptotic
formula (\ref{E:finnys}) in Theorem \ref{T:mixx}.

\begin{remark}
\label{R:KL} \textrm{There exist explicit formulas for the Karhunen-Lo\`{e}%
ve characteristics of various Gaussian processes. For Brownian motion,
Brownian bridge, and OU processes, such formulas can be found in \cite{CP}.
For OU bridges, one can consult \cite{DM0,Co}, and for the Gaussian process
introduced in \cite{DM}, the Karhunen-Lo\`{e}ve decomposition can be found
in the same paper. Unfortunately, even for classical fractional Gaussian
processes, e.g., fBm or fOU, the Karhunen-Lo\`{e}ve characteristics are not
known. In \cite{Cor} (see also \cite{C}), Corlay developed a powerful
numerical method to approximate Karhunen-Lo\`{e}ve eigenvalues and
eigenfunctions. Corlay uses the Nystr\"{o}m method associated with the
trapezoidal integration rule combined with the Richardson-Romberg
extrapolation in his work. }
\end{remark}

The independence of $W$ and $X$ imply that the density $D_{T}$ of $S_{T}$ is
given by a mixing formula (\ref{E:sic2}) involving $\tilde{p}_{1}$ via the
self-similar scaling property $\widetilde{p}_{T}(y)=T^{-H}\widetilde{p}%
_{1}(T^{-H}y)$. A delicate use of Laplace's method then allows to translate
Theorem \ref{T:mixx} into small-$T$ asymptotics for $D_{T}\left( x\right) $
for any $x$ which is \textquotedblleft out of the money\textquotedblright\
in the context of call pricing, in the sense that the big $O$ term depends
on a parameter $\varepsilon >0$ to allow for $x>s_{0}+\varepsilon $ (future
stock price parameter $x$, which stands in for strike price $K$ when one
computes an IV, exceeds initial stock price $s_{0}$ by a margin $\varepsilon 
$). We find (Theorem \ref{T:charc}) that for all for $x>s_{0}+\varepsilon $ 
\begin{align}
& D_{T}(x)=\frac{\sqrt{s_{0}}}{2^{\frac{n_{1}(1)}{2}}\Gamma \left( \frac{%
n_{1}(1)}{2}\right) }\lambda _{1}(1)^{-\frac{n_{1}(1)}{4}}\prod_{k=2}^{%
\infty }\left( \frac{\lambda _{1}(1)}{\lambda _{1}(1)-\rho _{k}(1)}\right) ^{%
\frac{n_{k}}{2}}  \notag \\
& \quad \times x^{-\frac{3}{2}}\left( \log \frac{x}{s_{0}}\right) ^{\frac{%
n_{1}(1)-2}{2}}T^{-\frac{(2H+1)n_{1}(1)}{4}}\left( \frac{x}{s_{0}}\right) ^{-%
\frac{\sqrt{4+\lambda _{1}(1)T^{2H+1}}}{2\sqrt{\lambda _{1}(1)}T^{H+\frac{1}{%
2}}}}  \notag \\
& \quad \times \left( 1+O\left( T^{2H+1}\right) \right) \left(
1+O_{\varepsilon }\left( T^{\frac{2H+1}{4}}\left( \log \frac{x}{s_{0}}%
\right) ^{-\frac{1}{2}}\right) \right)  \label{DTconstant}
\end{align}%
where the repeated notation $\left( 1\right) $ refers to KL elements for $%
T=1 $, and where $n_{k}$ is the multiplicity of the $k$th largest KL
eigenvalue $\rho _{k}$. The symbol $O$ depends only on the covariance of $X$%
, but not on $x$ or $\varepsilon $. The symbol $O_{\varepsilon }$ depends on
the covariance of $X$ and on $\varepsilon $, but not $x$. We prove formula (%
\ref{DTconstant}) under the assumptions that $r=0$ and the volatility
process $X^{(H)}$ is centered. The case where $r>0$ and the process $X^{(H)}$
is noncentered is more complicated and will be addressed in future
publications.

Being able to establish the precise $x$-behavior of the error terms above is
crucial to tranposing the behavior of $D_{T}\left( x\right) $ to the
functions $C$ and $I$. Specifically, we obtain the following for the
out-of-the-money call as $T\rightarrow 0^{+}$ (Theorem \ref{T:call}) : for $%
K>s_{0}$,%
\begin{equation}
C(T)=MT^{\frac{(2H+1)(4-n_{1}(1))}{4}}\left( \frac{s_{0}}{K}\right)
^{\lambda _{1}(1)^{-\frac{1}{2}}T^{-H-\frac{1}{2}}}\left( 1+O\left( T^{\frac{%
2H+1}{4}}\right) \right)  \label{Cee}
\end{equation}%
where the big $O$ above does not depend on $K$ if it is away from $s_{0}$,
and the constant $M$ is explicit and proportional to the constant on the
right-hand side of line (\ref{DTconstant}). A nearly indentical result is
obtained for out-of-the-money put prices $P\left( T,K\right) $ (for $%
0<K<s_{0})$ using symmetries of the problem (Theorem \ref{T:put}).

Ultimately, relying on a general result of Gao and Lee \cite{GL} for
computing the small-time asymptotics of IV based on those of $C$, we obtain
in Theorems \ref{T:fiv} and \ref{T:fivy} that for $0<K\neq s_{0}$,%
\begin{equation}
I(T)=\frac{\lambda _{1}(1)^{\frac{1}{4}}\sqrt{\left\vert \log \frac{K}{s_{0}}%
\right\vert }}{\sqrt{2}}T^{\frac{2H-1}{4}}+O\left( T^{\frac{6H+1}{4}}\log 
\frac{1}{T}\right)  \label{Eye}
\end{equation}%
where the big $O$ is again uniform over $K$ in any compact interval away
from $0$ and $s_{0}$. The dominant factor in the expression (\ref{Cee}) for $%
C$, and its analogue for $P$, is the exponential one. In the expression (\ref%
{Eye}) for $I$, there is only one candidate for a dominant term.
Consequently, one gets a way to estimate $H$ using call or put prices or IVs
away from the money as empirical statistics:%
\begin{eqnarray*}
H &=&\lim_{T\rightarrow 0}\frac{\log \log \frac{1}{C(T,K)}}{\log \frac{1}{T}}%
-\frac{1}{2} \\
&=&\lim_{T\rightarrow 0}\frac{\log \log \frac{1}{P(T,K)}}{\log \frac{1}{T}}-%
\frac{1}{2} \\
&=&2\lim_{T\rightarrow 0}\frac{\log \frac{1}{I(T,K)}}{\log \frac{1}{T}}+%
\frac{1}{2}
\end{eqnarray*}%
where the first line holds for $K>s_{0}$, the second for $K<s_{0}$, and the
third holds for all $K\neq s_{0}$ (Corollaries \ref{C:corrola}, \ref%
{C:toyota}, \ref{C:fiv}, and \ref{C:fivy}.) These expressions for $H$ do not
depend on any of the model parameters and statistics, and are in this sense
model free within the class of self-similar models. However, in practice,
since the regime $T\rightarrow 0$ is limited by the ability to trade options
in a liquid way sufficiently close to maturity, the full asymptotics in (\ref%
{Cee}) and (\ref{Eye}) will typically be needed to help control the
estimation error.

We notice that the above asymptotics for $C$ and $I$ formally lose
information when $K=s_{0}$, since the expression $\left\vert \log \left(
K/s_{0}\right) \right\vert $ is zero and thus kills the dominant terms.
Hence the estimators for $H$ above are not longer valid in that case. We
investigate this at-the-money situation in some detail. The delicate
calculations are largely performed \textquotedblleft by
hand\textquotedblright . The resulting asymptotics seem to rely on model
statistics which cannot be related to the KL elements in any simple fashion,
since they require computing the moments $\mu _{1/2}$ and $\mu _{3/2}$ of
order $1/2$ and $3/2$ of the non-explicit integrated variance's law. As $%
T\rightarrow 0$, we get in Corollary \ref{C:bis} that%
\begin{equation*}
C(T,s_{0})=\frac{s_{0}\mu _{1/2}}{\sqrt{2\pi }}T^{H+\frac{1}{2}}-\frac{%
s_{0}\mu _{3/2}}{24\sqrt{2\pi }}T^{3H+\frac{3}{2}}+O\left( T^{5H+\frac{5}{2}%
}\right) ,
\end{equation*}%
and in Theorem \ref{T:aiv} that 
\begin{equation}
I(T,s_{0})=\mu _{1/2}T^{H}+\frac{\left( \mu _{1/2}\right) ^{3}-\mu _{3/2}}{24%
}T^{3H+1}+O\left( T^{5H+2}\right) .  \label{Vee}
\end{equation}%
Again, simple $H$-estimators can result, which do not rely on the moments $%
\mu _{1/2}$ and $\mu _{3/2}$, such as Theorem \ref{T:recover} :%
\begin{equation*}
H=\lim_{T\rightarrow 0}\frac{\log \frac{1}{I(T,s_{0})}}{\log \frac{1}{T}}.
\end{equation*}

To illustrate the usage of our various asymptotic formulas numerically, we
provide simulated stock prices, with corresponding call prices and IVs, from
the self-similar volatility model, using a classical a Monte-Carlo method.
Using market-realistic parameter choices, we show how close prices and IVs
are to our asymptotic formulas, noting that the fit is good in the call
price case, and is excellent in the IV case, for time-to-maturity as large
as 2 weeks. It is then not surprising when we show that our IV-based
model-free calibration formulas for $H$ are accurate to $2$ decimal points
up to 7 days in most cases, and 14 days in some cases. Being able to use the
longest-possible time to maturity is important in practice because of
liquidity considerations. This is all explained in Section \ref{NUM}.

\bigskip

The remainder of the article is structured as follows. Some mathematical
background on Gaussian volatility models, taken largely from \cite{GVZ}, is
in Section \ref{S:tfBm}. Scaling consequences of self-similarity for the
density of the integrated variance are provided in Section \ref{S:dva}.
Section \ref{S:asset} contains the main asymptotic analysis of $S_{T}$'s
density. Consequences for call, put, and IV asymptotics away from the money
are in Sections \ref{S:call} and \ref{S:gl} respectively. Sections \ref%
{S:atthe} and \ref{S:atmiv} contain call and IV asymptotics at the money.
The numerics in Section \ref{NUM} finish this paper.

\section{Mathematical background on Gaussian stochastic volatility models}

\label{S:tfBm} In the present section, we consider the Gaussian stochastic
volatility model defined by (\ref{E:svm}). Let us fix the time horizon $T> 0$%
, and denote by $m$ and $K$ the mean function and the covariance function of
the process $X$ given by $m(t)=\mathbb{E}[X_t]$, $t\in[0,T]$ and 
\begin{equation*}
K(t,s)=\mathbb{E}\left[\left(X_t-m(t)\right)\left(X_s-m(s)\right)\right]%
,\quad t\in[0,T]^2,
\end{equation*}
respectively. It will be assumed that $K(s,s)> 0$ if $0\le s\le T$.

The following formula is valid for the distribution density $D_t$ of the
asset price $S_t$ in the Gaussian model described by (\ref{E:svm}): 
\begin{align}
&D_t(x)=\frac{\sqrt{s_0e^{rt}}}{\sqrt{2\pi t}}x^{-\frac{3}{2}}  \notag \\
&\int_0^{\infty} y^{-1}\exp\left\{-\left[\frac{\log^2\frac{x}{s_0e^{rt}}}{%
2ty^2}+\frac{ty^2}{8}\right]\right\}\widetilde{p}_t(y)dy.  \label{E:sic2}
\end{align}
In (\ref{E:sic2}), $\widetilde{p}_t$ is the distribution density of the
random variable 
\begin{equation}
\widetilde{Y}_t=\left\{\frac{1}{t}\int_0^tX^2_sds\right\}^{\frac{1}{2}}.
\label{E:ddrv}
\end{equation}
The function $\widetilde{p}_t$ is called the mixing density (see \cite{G}).
The proof of formula (\ref{E:sic2}) can be found in \cite{GS,G}.

Applying the Karhunen-Lo\`{e}ve theorem to the Gaussian process $%
\left\{X_t\right\}_{t\in[0,T]}$, we obtain 
\begin{equation}
\widetilde{X}_t=\sum_{n=1}^{\infty}\sqrt{\lambda_n}e_n(t)Z_n.  \label{E:KL}
\end{equation}
In (\ref{E:KL}), $\{e_n=e_{n,T}\}$ is an orthonormal system of
eigenfunctions of the covariance operator 
\begin{equation*}
\mathcal{K}(f)(t)=\int_{0,T}f(s)K(t,s)ds,\quad f\in L^2[0,T],\quad 0\le t\le
T,
\end{equation*}
and $\{\lambda_n=\lambda_{n}(T)\}$, $n\ge 1$, are the corresponding
eigenvalues (counting the multiplicities). The symbols $Z_n=Z_{n,T}$, $n\ge
1 $, in (\ref{E:KL}) stand for a system of iid $\mathcal{N}(0,1)$ random
variables. We will always assume that the orthonormal system $\{e_n\}$ is
rearranged so that 
\begin{equation*}
\lambda_1=\lambda_2=\dots=\lambda_{n_1}>\lambda_{n_1+1}=\lambda_{n_1+2}
=\dots=\lambda_{n_1+n_2}>\dots
\end{equation*}
For the sake of shortness, we introduce the following notation: 
\begin{equation*}
\rho_1=\lambda_1,\,\rho_2=\lambda_{n_1+1},\,\rho_3=\lambda_{n_1+n_2+1},%
\cdots,
\end{equation*}
\begin{equation*}
\delta_n=\delta_{n}(T)=\int_0^Tm(t)e_n(t)dt,\quad n\ge 1,
\end{equation*}
\begin{equation*}
s=s(T)=\int_0^Tm(t)^2dt,\quad\delta=\delta(T)=\frac{1}{\lambda_1}%
\sum_{n=1}^{n_1}\delta_n^2.
\end{equation*}

The mixing density $\widetilde{p}_{T}$ is related to the density $p_{T}$ of
the integrated variance 
\begin{equation*}
Y_{T}=\int_{0}^{T}X_{t}^{2}dt
\end{equation*}%
as follows: 
\begin{equation}
\widetilde{p}_{T}(y)=2Typ_{T}\left( Ty^{2}\right) .  \label{E:sviaz}
\end{equation}

The next theorem, characterizing the asymptotic behavior of the density $p_T$%
, was established in \cite{GVZ}.

\begin{theorem}
\label{T:gella} If $\delta> 0$, then the following asymptotic formula holds: 
\begin{align}
p_T(x)&=Cx^{\frac{n_1-3}{4}}\exp\left\{\sqrt{\frac{\delta}{\lambda_1}}\sqrt{x%
}\right\} \exp\left\{-\frac{x}{2\lambda_1}\right\}  \notag \\
&\quad\times\left(1+O\left(x^{-\frac{1}{2}}\right)\right)  \label{E:finn}
\end{align}
as $x\rightarrow\infty$, where 
\begin{equation}
C=\frac{A}{2\sqrt{2\pi}}\lambda_1^{-\frac{1}{2}}\left(\sum_{n=1}^{n_1}%
\delta_n^2\right) ^{-\frac{n_1-1}{4}}\exp\left\{\frac{s-\sum_{n=1}^{\infty}%
\delta_n^2-\sum_{n=1}^{n_1}\delta_n^2} {2\lambda_1}\right\}.  \label{E:rr}
\end{equation}
The constant $A$ in (\ref{E:rr}) is given by 
\begin{equation*}
A=\prod_{k=2}^{\infty}\left(\frac{\lambda_1}{\lambda_1-\rho_k}\right)^{\frac{%
n_k}{2}} \exp\left\{\frac{1}{2}\sum_{k=2}^{\infty}\frac{1}{\lambda_1-\rho_k}%
\left(\sum_{n=n_1+\cdots+n_{k-1}+1}
^{n_1+\cdots+n_k}\delta_n^2\right)\right\}.
\end{equation*}

On the other hand, for a centered Gaussian process $X$, we have 
\begin{align}
p_T(x)&=Cx^{\frac{n_1-2}{2}}\exp\left\{-\frac{x}{2\lambda_1}\right\}
\left(1+O\left(x^{-\frac{1}{2}}\right)\right)  \label{E:finna}
\end{align}
as $x\rightarrow\infty$, where 
\begin{equation}
C=\frac{1}{2^{\frac{n_1}{2}}\Gamma\left(\frac{n_1}{2}\right)\lambda_1^{\frac{%
n_1}{2}}} \prod_{k=2}^{\infty}\left(\frac{\lambda_1}{\lambda_1-\rho_k}%
\right)^{\frac{n_k}{2}}.  \label{E:ho}
\end{equation}
\end{theorem}

The next assertion follows form Theorem \ref{T:gella}.

\begin{corollary}
\label{C:coriol} The following are true:

\begin{enumerate}
\item If $n_1=1$, then 
\begin{align}
p_T(x)=&C x^{-\frac{1}{2}}\exp\left\{\frac{\delta_1}{\lambda_1}\sqrt{x}%
\right\} \exp\left\{-\frac{x}{2\lambda_1}\right\}  \notag \\
&\quad\times\left(1+O\left(x^{-\frac{1}{2}}\right)\right)  \label{E:finnnl}
\end{align}
as $x\rightarrow\infty$, where $C$ is given by (\ref{E:rr}).

\item Suppose $X$ is a centered Gaussian process with $n_1=1$. Then 
\begin{equation}
p_T(x)=C x^{-\frac{1}{2}}\exp\left\{-\frac{x}{2\lambda_1}\right\}
\left(1+O\left(x^{-\frac{1}{2}}\right)\right)  \label{E:finanl}
\end{equation}
as $x\rightarrow\infty$.
\end{enumerate}
\end{corollary}

It was established in \cite{GVZ} that Gaussian stochastic volatility models
are risk-neutral.

\begin{lemma}
\label{L:environ} In the Gaussian stochastic volatility model, the
discounted asset price process $t\mapsto e^{-rt}S_t$ is a $\{\mathcal{F}_t\}$%
-martingale.
\end{lemma}

\section{Fractional Gaussian stochastic volatility models}

\label{S:dva} The paper \cite{GVZ} is mostly devoted to the extreme strike
asymptotics of option pricing functions and the implied volatility in
Gaussian stochastic volatility models. The present paper deals with Gaussian
models, in which the volatility process is self-similar, and also with
small-time asymptotic behavior of option pricing functions and the implied
volatility in such models.

\begin{definition}
\label{D:self} Let $0< H< 1$. A stochastic process $X^{(H)}$ is called $H$%
-self-similar if for every $a> 0$, $X^{(H)}_{at}\overset{d}{=}a^HX^{(H)}_t$.
Here $\overset{d}{=}$ means the equality of all finite-dimensional
distributions.
\end{definition}

It is easy to see that if the process $X^{(H)}$ is $H$-self-similar, then $%
X^{(H)}_0=0$. It will always be assumed in the sequel that the self-similar
process $X^{(H)}$ is stochastically continuous. For a Gaussian process $X$,
the $H$-self-similarity condition is expressed in terms of the covariance
function $C$ as follows: 
\begin{equation*}
C(at,as)=a^{2H}C(t,s),\quad(t,s)\in[0,T]^2.
\end{equation*}
We refer the interested reader to \cite{EM,T} for more information on
self-similar stochastic processes.

Let us consider the following Gaussian stochastic volatility model: 
\begin{equation}
dS_t=rS_tdt+|X_t^{(H)}|S_tdW_t,\quad S_0=s_0,  \label{E:1}
\end{equation}
where $s_0> 0$ is the initial condition for the asset price process $S$, $W$
is a standard Brownian motion, and $X^{(H)}$ is a continuous $H$%
-self-similar adapted Gaussian process. The process $S$ characterizes the
dynamics of the asset price in the stochastic volatility model, where the
volatility is desribed by the absolute value of a self-similar Gaussian
process. It will be assumed throughout the paper that the model in (\ref{E:1}%
) is uncorrelated, which means that the processes $X^{(H)}$ and $W$ are
independent. We will often suppress the parameter $H$ in various symbols
used in the paper. A popular example of a self-similar Gaussian process is
fractional Brownian motion $B^{(H)}$ (see, e.g., \cite{No}). Note that
fractional Brownian motion is the only process that is non-trivial,
self-similar, Gaussian, and has stationary increments.

Exactly as in Section \ref{S:tfBm}, we will denote by $p_t$ the denstiy of
the integrated variance, 
\begin{equation*}
Y_t=\int_0^t\left(X_s^{(H)}\right)^2ds,
\end{equation*}
and by $\widetilde{p}_t$ the density of the random variable 
\begin{equation*}
\widetilde{Y}_t=\left[\frac{1}{t}\int_0^t\left(X_s^{(H)}\right)^2ds\right]^{%
\frac{1}{2}}
\end{equation*}
(the mixing density). Since the process $X^{(H)}$ is self-similar, we have $%
Y_{at}=a^{2H+1}Y_t. $ Moreover, the following equality holds: $\mathbb{P}%
\left(Y_t> y\right)=\mathbb{P}\left(Y_1> t^{-2H-1}y\right), $ and hence, 
\begin{equation}
p_t(y)=t^{-2H-1}p_1\left(t^{-2H-1}y\right).  \label{E:5}
\end{equation}

The next assertion characterizes the small-time asymptotics of the mixing
density.

\begin{theorem}
\label{T:mixx} (i)\,\,For every $x> 0$, the following asymptotic formula
holds for the mixing density $\widetilde{p}_T$ in the model discribed by (%
\ref{E:1}): 
\begin{align}
\widetilde{p}_T(x)&=2CT^{-\frac{H(n_1(1)+1)}{2}}x^{\frac{n_1(1)-1}{2}}
\exp\left\{\sqrt{\frac{\delta(1)}{\lambda_1(1)}}\frac{x}{T^H}\right\}
\exp\left\{-\frac{x^2}{2T^{2H}\lambda_1(1)}\right\}  \notag \\
&\quad\times\left(1+O_x\left(T^H\right)\right)  \label{E:finnys}
\end{align}
as $T\rightarrow 0$, where 
\begin{align}
C&=\frac{A}{2\sqrt{2\pi}}\lambda_1(1)^{-\frac{1}{2}}\left(%
\sum_{n=1}^{n_1(1)}\delta_n(1)^2\right) ^{-\frac{n_1(1)-1}{4}}  \notag \\
&\quad\times\exp\left\{\frac{s(1)-\sum_{n=1}^{\infty}\delta_n(1)^2-%
\sum_{n=1}^{n_1}\delta_n(1)^2} {2\lambda_1(1)}\right\},  \label{E:rro}
\end{align}
and the constant $A$ in (\ref{E:rro}) is given by 
\begin{align*}
A&=\prod_{k=2}^{\infty}\left(\frac{\lambda_1(1)}{\lambda_1(1)-\rho_k(1)}%
\right)^{\frac{n_k(1)}{2}} \\
&\quad\exp\left\{\frac{1}{2}\sum_{k=2}^{\infty}\frac{1}{\lambda_1(1)-%
\rho_k(1)}\left(\sum_{n=n_1(1)+\cdots+n_{k-1}(1)+1}
^{n_1(1)+\cdots+n_k(1)}\delta_n(1)^2\right)\right\}.
\end{align*}
\newline
\newline
(ii)\,\,If the process $X^{(H)}$ is centered, then 
\begin{align}
\widetilde{p}_T(x)&=2CT^{-Hn_1(1)}x^{n_1(1)-1} \exp\left\{-\frac{x^2}{%
2t^{2H}\lambda_1(1)}\right\}  \notag \\
&\quad\left(1+O_x\left(T^H\right) \right)  \label{E:x2}
\end{align}
as $T\rightarrow 0$, where 
\begin{equation}
C=\frac{1}{2^{\frac{n_1(1)}{2}}\Gamma\left(\frac{n_1(1)}{2}\right)}%
\lambda_1^{-\frac{n_1(1)}{2}}\prod_{k=2}^{\infty}\left(\frac{\lambda_1(1)}{%
\lambda_1(1)-\rho_k(1)}\right)^ {\frac{n_k(1)}{2}}.  \label{E:fig}
\end{equation}
\newline
\newline
(iii)\,\,If the process $X^{(H)}$ is centered and $n_1(1)=1$, then 
\begin{equation}
\widetilde{p}_T(x)=2CT^{-H}\exp\left\{-\frac{x^2}{2T^{2H}\lambda_1(1)}%
\right\}\left(1+O_x\left(T^H\right) \right)  \label{E:x}
\end{equation}
as $T\rightarrow 0$, where the constant $C$ is given by (\ref{E:fig}) with $%
n_1(1)=1$.
\end{theorem}

Proof. It follows from (\ref{E:sviaz}) and (\ref{E:5}) that 
\begin{equation}
\widetilde{p}_T(x)=2T^{-2H}xp_1\left(T^{-2H}x^2\right).  \label{E:h}
\end{equation}
Since $X^{(H)}$ is a Gaussian process, we can use formula (\ref{E:finn}).
This gives 
\begin{align}
p_1(x)&=Cx^{\frac{n_1(1)-3}{4}}\exp\left\{\sqrt{\frac{\delta(1)}{\lambda_1(1)%
}}\sqrt{x}\right\} \exp\left\{-\frac{x}{2\lambda_1(1)}\right\}  \notag \\
&\quad\times\left(1+O\left(x^{-\frac{1}{2}}\right)\right)  \label{E:finny}
\end{align}
as $x\rightarrow\infty$, where the constant $C$ is given by (\ref{E:rro}).
If the process $X^{(H)}$ is centered, then formulas (\ref{E:finna}) and (\ref%
{E:ho}) imply that 
\begin{equation}
p_1(x)=Cx^{\frac{n_1(1)-2}{2}}\exp\left\{-\frac{x}{2\lambda_1(1)}\right\}
\left(1+O\left(x^{-\frac{1}{2}}\right)\right)  \label{E:finanll}
\end{equation}
as $x\rightarrow\infty$, where the constant $C$ is given by (\ref{E:fig}).
Now, Theorem \ref{T:mixx} can be derived from from (\ref{E:h}), (\ref%
{E:finny}), and (\ref{E:finanll}).

\section{Small-time asymptotics of the asset price density in self-similar
Gaussian stochastic volatility models with centered volatility.}

\label{S:asset} In this section, we restrict ourselves to the case where the
process $X^{(H)}$ is an adapted continuous $H$-self-similar centered
Gaussian process. Recall that we assume $r=0$.

Of our interest in the present paper are asymptotic estimates of the density 
$D_T(x)$ as $T\rightarrow 0$, which are uniform with respect to the values
of $x> 0$ separated from $s_0$ (away-from-the-money regime). Here we
distinguish among two special cases. In the first case, we fix $\varepsilon>
0$, and consider asymptotic expansions as $t\rightarrow 0$, which are unform
with respect to $x> s_0+\varepsilon$. The notation $O_{\varepsilon}(%
\phi(t,x))$ as $t\rightarrow 0$, where $\phi$ is a positive function of two
variables, means that the $O$-large estimate holds as $t\rightarrow 0$
uniformly with respect to $x> s_0+\varepsilon$. In the second case, we fix $%
\varepsilon$ with $0<\varepsilon< s_0$, and assume that $0< x<
s_0-\varepsilon$. The same notation $O_{\varepsilon}(\phi(t,x))$ will be
used in the second case.

Since $\widetilde{p}_T(y)=T^{-H}\widetilde{p}_1(T^{-H}y)$, formula (\ref%
{E:sic2}) implies that 
\begin{align}
&D_T(x)=\frac{\sqrt{s_0}}{\sqrt{2\pi}}T^{-H-\frac{1}{2}}x^{-\frac{3}{2}} 
\notag \\
&\quad\times\int_0^{\infty} y^{-1}\exp\left\{-\left[\frac{\log^2\frac{x}{s_0}%
}{2Ty^2}+\frac{Ty^2}{8}\right]\right\} \widetilde{p}_1(T^{-H}y)dy  \notag \\
&=\frac{\sqrt{s_0}}{\sqrt{2\pi}}T^{-H-\frac{1}{2}}x^{-\frac{3}{2}}  \notag \\
&\quad\times\int_0^{\infty} u^{-1}\exp\left\{-\left[\frac{\log^2\frac{x}{s_0}%
}{2T^{2H+1}u^2}+\frac{T^{2H+1}u^2}{8}\right]\right\} \widetilde{p}_1(u)du.
\label{E:sictr}
\end{align}

The next assertion is one of the main results of the present paper. It
characterizes the small-time asymptotic behavior of the asset price density
in a Gaussian model with a centered self-similar volatility process.

\begin{theorem}
\label{T:charc} Fix $\varepsilon> 0$ and let $x> s_0+\varepsilon$. Then as $%
T\rightarrow 0$, the following asymptotic formula holds for the asset price
density $D_T$ in the model described by (\ref{E:1}): 
\begin{align}
&D_T(x)=\frac{\sqrt{s_0}}{2^{\frac{n_1(1)}{2}}\Gamma\left(\frac{n_1(1)}{2}%
\right)}\lambda_1(1)^{-\frac{n_1(1)}{4}} \prod_{k=2}^{\infty}\left(\frac{%
\lambda_1(1)}{\lambda_1(1)-\rho_k(1)}\right) ^{\frac{n_k}{2}}x^{-\frac{3}{2}}
\notag \\
&\quad\times\left(\log\frac{x}{s_0}\right)^{\frac{n_1(1)-2}{2}}T^{-\frac{%
(2H+1)n_1(1)}{4}} \left(\frac{x}{s_0}\right)^{-\frac{\sqrt{%
4+\lambda_1(1)T^{2H+1}}} {2\sqrt{\lambda_1(1)}T^{H+\frac{1}{2}}}}  \notag \\
&\quad\times\left(1+O\left(T^{2H+1}\right)\right)\left(1+O_{\varepsilon}%
\left(T^{\frac{2H+1}{4}} \left(\log\frac{x}{s_0}\right)^{-\frac{1}{2}%
}\right)\right).  \label{E:forr}
\end{align}
\end{theorem}

Proof. Fix $x> 0$, and denote 
\begin{equation}
J_x(T)=\int_0^{\infty} u^{-1}\exp\left\{-\left[\frac{\log^2\frac{x}{s_0}}{%
2T^{2H+1}u^2}+\frac{T^{2H+1}u^2}{8}\right]\right\} \widetilde{p}_1(u)du
\label{E:esti}
\end{equation}
It is clear from (\ref{E:sictr}) that the small-time asymptotic behavior of
the density $D_T(x)$ is determined by that of the integral $J_x(T)$.

The next lemma will allow us to use Theorem \ref{T:mixx} to estimate the
integral in (\ref{E:esti}).

\begin{lemma}
\label{L:estiv} Fix $\alpha\in\mathbb{R}$, $b> 0$, and $\varepsilon> 0$. Let 
$x> s_0+\varepsilon$, and suppose $f$ is an integrable function on $[0,b]$.
Then 
\begin{align*}
&\int_0^b u^{\alpha}\exp\left\{-\left[\frac{\log^2\frac{x}{s_0}}{2T^{2H+1}u^2%
}+\frac{T^{2H+1}u^2}{8}\right]\right\} |f(u)|du \\
&=O_{\varepsilon}\left(\exp\left\{-\frac{\log^2\frac{x}{s_0}}{2b^2T^{2H+1}}%
\right\}\right)
\end{align*}
as $t\rightarrow 0$.
\end{lemma}

Proof. The lemma is trivial if $\alpha\ge 0$. For $\alpha< 0$, we have 
\begin{align}
&\int_0^b u^{\alpha}\exp\left\{-\left[\frac{\log^2\frac{x}{s_0}}{2T^{2H+1}u^2%
}+\frac{T^{2H+1}u^2}{8}\right]\right\} |f(u)|du  \notag \\
&\le\int_0^b u^{\alpha}\exp\left\{-\frac{\log^2\frac{x}{s_0}}{2T^{2H+1}u^2}%
\right\}|f(u)|du.  \label{E:al}
\end{align}
The following equality holds for every $A> 0$: 
\begin{equation*}
\left(u^{\alpha}\exp\left\{-\frac{A}{u^2}\right\}\right)^{\prime}=\left[%
2Au^{\alpha-3}+\alpha u^{\alpha-1} \right] \exp\left\{-\frac{A}{u^2}\right\}.
\end{equation*}
It follows that for $2A>-\alpha b^2$, the function 
\begin{equation*}
u\mapsto\frac{1}{u^{\alpha}}\exp\left\{-\frac{A}{u^2}\right\}
\end{equation*}
is increasing on the interval $(0,b]$. Set 
\begin{equation*}
A=\frac{\log^2\frac{x}{s_0}}{2T^{2H+1}}.
\end{equation*}
Using (\ref{E:al}), we obtain 
\begin{align}
&\int_0^b u^{\alpha}\exp\left\{-\left[\frac{\log^2\frac{x}{s_0}}{2T^{2H+1}u^2%
}+\frac{T^{2H+1}u^2}{8}\right]\right\} |f(u)|du  \notag \\
&\le b^{\alpha}\exp\left\{-\frac{\log^2\frac{x}{s_0}}{2b^2T^{2H+1}}%
\right\}\int_0^b|f(u)|du,  \label{E:al1}
\end{align}
provided that $\log^2\frac{x}{s_0}> b^2T^{2H+1}$. It is clear that the
previous inequality holds for small enough values of $T$ provided that $x>
s_0+\varepsilon$.

Finally, Lemma \ref{L:estiv} follows from (\ref{E:al1}).

Using (\ref{E:h}) and (\ref{E:finanll}), we obtain 
\begin{equation}
\widetilde{p}_1(y)=\widetilde{A}y^{n_1(1)-1}\exp\left\{-\frac{y^2}{%
2\lambda_1(1)}\right\} \left(1+O\left(y^{-1}\right) \right)  \label{E:ot1}
\end{equation}
as $y\rightarrow\infty$, where 
\begin{equation}
\widetilde{A}=\frac{2^{1-\frac{n_1(1)}{2}}}{\Gamma\left(\frac{n_1(1)}{2}%
\right)} \lambda_1^{-\frac{n_1(1)}{2}}\prod_{k=2}^{\infty}\left(\frac{%
\lambda_1(1)}{\lambda_1(1)-\rho_k(1)}\right) ^{\frac{n_k}{2}}.  \label{E:ot2}
\end{equation}
It is not hard to see that Lemma \ref{L:estiv} allows us to replace the
function $\widetilde{p}_1(u)$ in (\ref{E:esti}) by its approximation from (%
\ref{E:ot1}). This gives the following: 
\begin{align}
&J_x(T)=\widetilde{A}\int_0^{\infty}u^{n_1(1)-2}  \notag \\
&\exp\left\{-\left[\frac{\log^2\frac{x}{s_0}}{2T^{2H+1}u^2}+\left(\frac{%
T^{2H+1}}{8}+\frac{1}{2\lambda_1(1)} \right)u^2\right]\right\}
\left(1+O\left(u^{-1}\right)\right)du  \notag \\
&+O_{\varepsilon}\left(\exp\left\{-\frac{\log^2\frac{x}{s_0}}{2T^{2H+1}}%
\right\}\right)  \label{E:11}
\end{align}
as $T\rightarrow 0$.

To study the asymptotics of the function $t\mapsto J_{x}(T)$ defined by (\ref%
{E:11}), we consider the following two integrals: 
\begin{align}
& \widetilde{J}_{x}(T)=\widetilde{A}\int_{0}^{\infty }u^{n_{1}(1)-2}  \notag
\\
& \exp \left\{ -\left[ \frac{\log ^{2}\frac{x}{s_{0}}}{2T^{2H+1}u^{2}}%
+\left( \frac{T^{2H+1}}{8}+\frac{1}{2\lambda _{1}(1)}\right) u^{2}\right]
\right\} du  \label{E:12}
\end{align}%
and 
\begin{align}
& \widehat{J}_{x}(T)=\widetilde{A}\int_{0}^{\infty }u^{n_{1}(1)-3}  \notag \\
& \exp \left\{ -\left[ \frac{\log ^{2}\frac{x}{s_{0}}}{2T^{2H+1}u^{2}}%
+\left( \frac{T^{2H+1}}{8}+\frac{1}{2\lambda _{1}(1)}\right) u^{2}\right]
\right\} du.  \label{E:13}
\end{align}%
Set 
\begin{equation*}
\beta _{T}=\frac{\log ^{2}\frac{x}{s_{0}}}{2T^{2H+1}},\quad \gamma _{T}=%
\frac{T^{2H+1}}{8}+\frac{1}{2\lambda _{1}(1)}.
\end{equation*}%
Note that $\beta _{T}$ depends on $x$, while $\gamma _{T}$ does not. Then we
have 
\begin{equation*}
\widetilde{J}_{x}(T)=\widetilde{A}\int_{0}^{\infty }u^{n_{1}(1)-2}\exp
\left\{ -\left[ \frac{\beta _{T}}{u^{2}}+\gamma _{T}u^{2}\right] \right\} du
\end{equation*}%
and 
\begin{equation*}
\widehat{J}_{x}(T)=\widetilde{A}\int_{0}^{\infty }u^{n_{1}(1)-3}\exp \left\{
-\left[ \frac{\beta _{T}}{u^{2}}+\gamma _{T}u^{2}\right] \right\} du.
\end{equation*}%
Next, making a substitution 
\begin{equation*}
u=\left( \frac{\beta _{T}}{\gamma _{T}}\right) ^{\frac{1}{4}}v,
\end{equation*}%
we transform the previous integrals as follows: 
\begin{equation*}
\widetilde{J}_{x}(T)=\widetilde{A}\left( \frac{\beta _{T}}{\gamma _{T}}%
\right) ^{\frac{n_{1}(1)-1}{4}}\int_{0}^{\infty }v^{n_{1}(1)-2}\exp \left\{ -%
\sqrt{\beta _{T}\gamma _{T}}\left[ \frac{1}{v^{2}}+v^{2}\right] \right\} dv
\end{equation*}%
and 
\begin{equation*}
\widehat{J}_{x}(T)=\widetilde{A}\left( \frac{\beta _{T}}{\gamma _{T}}\right)
^{\frac{n_{1}(1)-2}{4}}\int_{0}^{\infty }v^{n_{1}(1)-3}\exp \left\{ -\sqrt{%
\beta _{T}\gamma _{T}}\left[ \frac{1}{v^{2}}+v^{2}\right] \right\} dv.
\end{equation*}%
Let us denote 
\begin{equation}
z(T)=\frac{1}{4}\sqrt{\frac{\lambda _{1}(1)T^{2H+1}+4}{\lambda
_{1}(1)T^{2H+1}}}.  \label{E:red}
\end{equation}%
Then we have 
\begin{equation}
\sqrt{\beta _{T}\gamma _{T}}=z(T)\left\vert \log \frac{x}{s_{0}}\right\vert .
\label{E:red0}
\end{equation}%
Therefore, 
\begin{align}
\widetilde{J}_{x}(T)& =\widetilde{A}\left( \frac{\beta _{T}}{\gamma _{T}}%
\right) ^{\frac{n_{1}(1)-1}{4}}  \notag \\
& \quad \times \int_{0}^{\infty }v^{n_{1}(1)-2}\exp \left\{ -z(T)\left\vert
\log \frac{x}{s_{0}}\right\vert \left[ \frac{1}{v^{2}}+v^{2}\right] \right\}
dv  \label{E:red1}
\end{align}%
and 
\begin{align}
\widehat{J}_{x}(T)& =\widetilde{A}\left( \frac{\beta _{T}}{\gamma _{T}}%
\right) ^{\frac{n_{1}(1)-2}{4}}  \notag \\
& \quad \times \int_{0}^{\infty }v^{n_{1}(1)-3}\exp \left\{ -z(T)\left\vert
\log \frac{x}{s_{0}}\right\vert \left[ \frac{1}{v^{2}}+v^{2}\right] \right\}
dv.  \label{E:red2}
\end{align}

It follows from (\ref{E:red}) that $z(T)\rightarrow\infty$ as $T\rightarrow
0 $. Our next goal is to apply Laplace's method to study the asymptotic
behavior of the functions $T\mapsto\widetilde{J}_x(T)$ and $T\mapsto\widehat{%
J}_x(T)$ as $T\rightarrow 0$. Note that the unique critical point of the
function $\psi(v)=v^{-2}+v^2$ is at $v=1$. Moreover, we have $%
\psi^{\prime\prime}(1)=8> 0$.

We will first reduce the integrals in (\ref{E:red1}) and (\ref{E:red2}) to
the integrals over the interval $[0,2]$ and give an error estimate. This
next assertion will be helpful.

\begin{lemma}
\label{L:red} Suppose $a\in\mathbb{R}$ and $0<\varepsilon< s_0$. Then 
\begin{align*}
&\int_2^{\infty}v^a \exp\left\{-\sqrt{\beta_T\gamma_T}\left[\frac{1}{v^2}+v^2%
\right]\right\}dv =O_{\varepsilon}\left(\exp\left\{-3\sqrt{\beta_T\gamma_T}%
\right\}\right)
\end{align*}
as $t\rightarrow 0$.
\end{lemma}

Proof. Fix a small number $r> 0$. Then for $0< T< T_0$, we have 
\begin{align*}
&\int_2^{\infty}v^a \exp\left\{-\sqrt{\beta_T\gamma_T}\left[\frac{1}{v^2}+v^2%
\right]\right\}dv \le\int_2^{\infty}v^a \exp\left\{-\sqrt{\beta_T\gamma_T}%
v^2\right\}dv \\
&\le c_r\int_2^{\infty} \exp\left\{-\left(\sqrt{\beta_T\gamma_T}%
-r\right)v^2\right\}dv \\
&=c_r\left(\sqrt{\beta_T\gamma_T}-r\right)^{-\frac{1}{2}} \int_{2\sqrt{\sqrt{%
\beta_T\gamma_T}-r}}^{\infty}e^{-u^2}du \le\widetilde{c}_r\exp\left\{-4\left(%
\sqrt{\beta_T\gamma_T}-r\right)\right\}.
\end{align*}

The proof of Lemma \ref{L:red} is thus completed.

Now, we are ready to apply Laplace's method to the integrals in (\ref{E:red1}%
) and (\ref{E:red2}). The dependence of the parameter $x$ in (\ref{E:red1})
and (\ref{E:red2}) is very simple. This allows us to obtain uniform error
estimates. By taking into account Lemma \ref{L:red}, we see that for every $%
\varepsilon >0$ and all $x>s_{0}+\varepsilon $, 
\begin{align}
& \widetilde{J}_{x}(T)=\frac{\widetilde{A}\sqrt{\pi }}{2}\left( \frac{\beta
_{T}}{\gamma _{T}}\right) ^{\frac{n_{1}(1)-1}{4}}\left( z(T)\left\vert \log 
\frac{x}{s_{0}}\right\vert \right) ^{-\frac{1}{2}}\exp \left\{
-2z(T)\left\vert \log \frac{x}{s_{0}}\right\vert \right\}  \notag \\
& \quad \left( 1+O_{\varepsilon }\left( \frac{1}{z(T)\left\vert \log \frac{x%
}{s_{0}}\right\vert }\right) \right) +O_{\varepsilon }\left( \exp \left\{
-3z(T)\left\vert \log \frac{x}{s_{0}}\right\vert \right\} \right)
\label{E:u2}
\end{align}%
and 
\begin{align}
& \widehat{J}_{x}(T)=\frac{\widetilde{A}\sqrt{\pi }}{2}\left( \frac{\beta
_{T}}{\gamma _{T}}\right) ^{\frac{n_{1}(1)-2}{4}}\left( z(T)\left\vert \log 
\frac{x}{s_{0}}\right\vert \right) ^{-\frac{1}{2}}\exp \left\{
-2z(T)\left\vert \log \frac{x}{s_{0}}\right\vert \right\}  \notag \\
& \quad \left( 1+O_{\varepsilon }\left( \frac{1}{z(T)\left\vert \log \frac{x%
}{s_{0}}\right\vert }\right) \right) +O_{\varepsilon }\left( \exp \left\{
-3z(T)\left\vert \log \frac{x}{s_{0}}\right\vert \right\} \right)
\label{E:u3}
\end{align}%
as $T\rightarrow 0$. Recall that the $O_{\varepsilon }$ estimates in (\ref%
{E:u2}) and (\ref{E:u3}) are uniform with respect to $x>s_{0}+\varepsilon $.
Since 
\begin{equation*}
J_{x}(T)=\widetilde{J}_{x}(T)+O_{\varepsilon }\left( \widehat{J}%
_{x}(T)\right) +O_{\varepsilon }\left( \exp \left\{ -\frac{\log ^{2}\frac{x}{%
s_{0}}}{2T^{2H+1}}\right\} \right) ,
\end{equation*}%
as $T\rightarrow 0$, formulas (\ref{E:u2}) and (\ref{E:u3}) imply that 
\begin{align*}
& J_{x}(T)=\frac{\widetilde{A}\sqrt{\pi }}{2}\left( \frac{\beta _{T}}{\gamma
_{T}}\right) ^{\frac{n_{1}(1)-1}{4}}\left( z(T)\left\vert \log \frac{x}{s_{0}%
}\right\vert \right) ^{-\frac{1}{2}}\exp \left\{ -2z(T)\left\vert \log \frac{%
x}{s_{0}}\right\vert \right\} \\
& \left( 1+\left( \frac{\beta _{T}}{\gamma _{T}}\right) ^{-\frac{1}{4}%
}\right) \left( 1+O_{\varepsilon }\left( \frac{1}{z(T)\left\vert \log \frac{x%
}{s_{0}}\right\vert }\right) \right) \\
& \quad +O_{\varepsilon }\left( \exp \left\{ -\frac{\log ^{2}\frac{x}{s_{0}}%
}{2T^{2H+1}}\right\} \right) +O_{\varepsilon }\left( \exp \left\{
-3z(T)\left\vert \log \frac{x}{s_{0}}\right\vert \right\} \right)
\end{align*}%
as $T\rightarrow 0$. Since for $T<1$, 
\begin{equation}
\frac{1}{4}\sqrt{\frac{\lambda _{1}(1)+4}{\lambda _{1}(1)}}T^{-H-\frac{1}{2}%
}>z(T)>\frac{1}{2}\lambda _{1}(1)^{-\frac{1}{2}}T^{-H-\frac{1}{2}},
\label{E:eqq}
\end{equation}%
we have 
\begin{align*}
& O_{\varepsilon }\left( \exp \left\{ -\frac{\log ^{2}\frac{x}{s_{0}}}{%
2T^{2H+1}}\right\} \right) +O_{\varepsilon }\left( \exp \left\{
-3z(T)\left\vert \log \frac{x}{s_{0}}\right\vert \right\} \right) \\
& =O_{\varepsilon }\left( \exp \left\{ -3z(T)\left\vert \log \frac{x}{s_{0}}%
\right\vert \right\} \right) \\
& =O_{\varepsilon }\left( \exp \left\{ -\frac{3}{2}\lambda _{1}(1)^{-\frac{1%
}{2}}T^{-H-\frac{1}{2}}\left\vert \log \frac{x}{s_{0}}\right\vert \right\}
\right)
\end{align*}%
as $T\rightarrow 0$, and therefore, 
\begin{align*}
& J_{x}(T)=\frac{\widetilde{A}\sqrt{\pi }}{2}\left( \frac{\beta _{T}}{\gamma
_{T}}\right) ^{\frac{n_{1}(1)-1}{4}}\left( z(T)\left\vert \log \frac{x}{s_{0}%
}\right\vert \right) ^{-\frac{1}{2}}\exp \left\{ -2z(T)\left\vert \log \frac{%
x}{s_{0}}\right\vert \right\} \\
& \left( 1+\left( \frac{\beta _{T}}{\gamma _{T}}\right) ^{-\frac{1}{4}%
}\right) \left( 1+O_{\varepsilon }\left( \frac{1}{z(T)\left\vert \log \frac{x%
}{s_{0}}\right\vert }\right) \right) \\
& +O_{\varepsilon }\left( \exp \left\{ -\frac{3}{2}\lambda _{1}(1)^{-\frac{1%
}{2}}T^{-H-\frac{1}{2}}\left\vert \log \frac{x}{s_{0}}\right\vert \right\}
\right)
\end{align*}%
as $T\rightarrow 0$. Moreover, for all $T<1$ and $x>s_{0}+\varepsilon $, 
\begin{equation*}
\left( \frac{\beta _{T}}{\gamma _{T}}\right) ^{-\frac{1}{4}}\geq c_{1}\frac{%
T^{\frac{2H+1}{4}}}{\sqrt{\left\vert \log \frac{x}{s_{0}}\right\vert }}\geq
c_{2}\frac{T^{\frac{2H+1}{2}}}{\left\vert \log \frac{x}{s_{0}}\right\vert }%
\geq c_{3}\frac{1}{z(T)\left\vert \log \frac{x}{s_{0}}\right\vert },
\end{equation*}%
and hence 
\begin{align*}
& \left( 1+\left( \frac{\beta _{T}}{\gamma _{T}}\right) ^{-\frac{1}{4}%
}\right) \left( 1+O_{\varepsilon }\left( \frac{1}{z(T)\left\vert \log \frac{x%
}{s_{0}}\right\vert }\right) \right) \\
& =\left( 1+O_{\varepsilon }\left( \left( \frac{\beta _{T}}{\gamma _{T}}%
\right) ^{-\frac{1}{4}}\right) \right) =\left( 1+O_{\varepsilon }\left( T^{%
\frac{2H+1}{4}}\left\vert \log \frac{x}{s_{0}}\right\vert ^{-\frac{1}{2}%
}\right) \right)
\end{align*}%
as $T\rightarrow 0$. Finally, 
\begin{align*}
J_{x}(T)& =\frac{\widetilde{A}\sqrt{\pi }}{2}\left( \frac{\beta _{T}}{\gamma
_{T}}\right) ^{\frac{n_{1}(1)-1}{4}}\left( z(T)\left\vert \log \frac{x}{s_{0}%
}\right\vert \right) ^{-\frac{1}{2}}\exp \left\{ -2z(T)\left\vert \log \frac{%
x}{s_{0}}\right\vert \right\} \\
& \quad \left( 1+O_{\varepsilon }\left( T^{\frac{2H+1}{4}}\left\vert \log 
\frac{x}{s_{0}}\right\vert ^{-\frac{1}{2}}\right) \right) \\
& \quad +O_{\varepsilon }\left( \exp \left\{ -\frac{3}{2}\lambda _{1}(1)^{-%
\frac{1}{2}}T^{-H-\frac{1}{2}}\left\vert \log \frac{x}{s_{0}}\right\vert
\right\} \right)
\end{align*}%
as $T\rightarrow 0$.

Recall that we assumed $r=0$. It follows from (\ref{E:sictr}) and (\ref%
{E:esti}) that 
\begin{align}
&D_T(x)=\frac{\sqrt{s_0}\widetilde{A}}{2\sqrt{2}}T^{-H-\frac{1}{2}}x^{-\frac{%
3}{2}}\left(\frac{\beta_T}{\gamma_T}\right)^{\frac{n_1(1)-1}{4}%
}\left(z(T)\left|\log\frac{x}{s_0}\right|\right)^{-\frac{1}{2}}  \notag \\
&\quad\times\exp\left\{-2z(T)\left|\log\frac{x}{s_0}\right|\right\}
\left(1+O_{\varepsilon}\left(T^{\frac{2H+1}{4}} \left|\log\frac{x}{s_0}%
\right|^{-\frac{1}{2}}\right)\right)  \notag \\
&\quad+O_{\varepsilon}\left(\exp\left\{-\frac{3}{2}\lambda_1(1)^{-\frac{1}{2}%
} T^{-H-\frac{1}{2}}\left|\log\frac{x}{s_0}\right|\right\}\right)
\label{E:u4}
\end{align}
as $T\rightarrow 0$.

Our next goal is to remove the last $O_{\varepsilon }$-term from formula (%
\ref{E:u4}). Analyzing the expressions in (\ref{E:u4}), we see that in order
to prove the statement formulated above, it suffices to show that there
exists a constant $c>0$ independent of $T<T_{0}$ and $x>s_{0}+\varepsilon $
and such that 
\begin{align}
\left( \frac{x}{s_{0}}\right) ^{-\frac{3}{2}\lambda _{1}(1)^{-\frac{1}{2}%
}T^{-H-\frac{1}{2}}}& \leq cT^{-H-\frac{1}{2}}x^{-\frac{3}{2}}\left( \log 
\frac{x}{s_{0}}\right) ^{\frac{n_{1}(1)-1}{4}}T^{-\frac{(2H+1)(n_{1}(1)-1)}{8%
}}T^{\frac{2H+1}{4}}  \notag \\
& \times \left( \log \frac{x}{s_{0}}\right) ^{-\frac{1}{2}}\left( \frac{x}{%
s_{0}}\right) ^{-2z(T)}T^{\frac{2H+1}{4}}\left( \log \frac{x}{s_{0}}\right)
^{-\frac{1}{2}}.  \label{E:bravo}
\end{align}%
The previous inequality is equivalent to the following: 
\begin{align}
\left( \frac{x}{s_{0}}\right) ^{-\frac{3}{2}\lambda _{1}(1)^{-\frac{1}{2}%
}T^{-H-\frac{1}{2}}}& \leq cT^{-\frac{(2H+1)(n_{1}(1)-1)}{8}}x^{-\frac{3}{2}%
}\left( \frac{x}{s_{0}}\right) ^{-2z(T)}  \notag \\
& \quad \times \left( \log \frac{x}{s_{0}}\right) ^{\frac{n_{1}(1)-1}{4}-1}
\label{E:nak-1}
\end{align}%
Since (\ref{E:eqq}) holds, the inequality in (\ref{E:nak-1}) follows from
the inequality 
\begin{align}
& \left( \frac{x}{s_{0}}\right) ^{-\frac{3}{2}\lambda _{1}(1)^{-\frac{1}{2}%
}T^{-H-\frac{1}{2}}}\leq cT^{-\frac{(2H+1)(n_{1}(1)-1)}{8}}  \notag \\
& \quad \times \left( \frac{x}{s_{0}}\right) ^{-\frac{3}{2}-\frac{1}{2}%
\lambda _{1}(1)^{-\frac{1}{2}}T^{-H-\frac{1}{2}}\sqrt{\lambda
_{1}(1)T^{2H+1}+4}}\left( \log \frac{x}{s_{0}}\right) ^{\frac{n_{1}(1)-1}{4}%
-1}.  \label{E:naks}
\end{align}%
To prove the inequality in (\ref{E:naks}), we observe that for every small
enough $\tau >0$ there exists a constant $c_{\tau ,\varepsilon }$ such that 
\begin{equation*}
c_{\tau ,\varepsilon }\left( \frac{x}{s_{0}}\right) ^{-\tau }\leq \left(
\log \frac{x}{s_{0}}\right) ^{\frac{n_{1}(1)-1}{4}-1}
\end{equation*}%
for all $x>s_{0}+\varepsilon $. Moreover, there exists $T_{\tau ,\varepsilon
}>0$ such that 
\begin{equation*}
\left( \frac{x_{0}}{s_{0}}\right) ^{-\tau T^{-H-\frac{1}{2}}}\leq \left( 
\frac{s_{0}+\varepsilon }{s_{0}}\right) ^{-\tau T^{-H-\frac{1}{2}}}\leq T^{-%
\frac{(2H+1)(n_{1}(1)-1)}{8}}
\end{equation*}%
for all $T<T_{\tau ,\varepsilon }$. Now, it is clear that (\ref{E:naks})
follows from the estimate 
\begin{align}
& \left( \frac{3}{2}\lambda _{1}(1)^{-\frac{1}{2}}-\tau \right) T^{-H-\frac{1%
}{2}}  \notag \\
& \geq \frac{3}{2}+\frac{1}{2}\lambda _{1}(1)^{-\frac{1}{2}}T^{-H-\frac{1}{2}%
}\sqrt{\lambda _{1}(1)T^{2H+1}+4}+\tau ,  \label{E:hor}
\end{align}%
for all $T<T_{\tau }$. It is not hard to see that there exist numbers $\tau $
and $T_{\tau }$, for which the inequality in (\ref{E:hor}) holds. This
establishes (\ref{E:bravo}), and it follows that 
\begin{align}
& D_{T}(x)=\frac{\sqrt{s_{0}}\widetilde{A}}{2\sqrt{2}}T^{-H-\frac{1}{2}}x^{-%
\frac{3}{2}}\left( \frac{\beta _{T}}{\gamma _{T}}\right) ^{\frac{n_{1}(1)-1}{%
4}}\left( z(T)\log \frac{x}{s_{0}}\right) ^{-\frac{1}{2}}  \notag \\
& \quad \times \exp \left\{ -2z(T)\log \frac{x}{s_{0}}\right\} \left(
1+O_{\varepsilon }\left( T^{\frac{2H+1}{4}}\left( \log \frac{x}{s_{0}}%
\right) ^{-\frac{1}{2}}\right) \right)  \label{E:uu4}
\end{align}%
as $T\rightarrow 0$, where $\widetilde{A}$ is given by (\ref{E:ot2}).
Formula (\ref{E:uu4}) will help us to characterize the asymptotic behavior
of the function $T\mapsto D_{T}(x)$.

Let us assume that $x> s_0+\varepsilon$. Then we have 
\begin{equation*}
\left(\frac{\beta_T}{\gamma_T}\right)^{\frac{n_1(1)-1}{4}}= \lambda_1(1)^{%
\frac{n_1(1)-1}{4}}\left(\log\frac{x}{s_0}\right)^{\frac{n_1(1)-1}{2}} T^{-%
\frac{(2H+1)(n_1(1)-1)}{4}}(1+h)^{-\frac{n_1(1)-1}{4}}
\end{equation*}
where $h=\frac{\lambda_1(1)T^{2H+1}}{4}$. Therefore, 
\begin{align}
\left(\frac{\beta_T}{\gamma_T}\right)^{\frac{n_1(1)-1}{4}}&= \lambda_1(1)^{%
\frac{n_1(1)-1}{4}}\left(\log\frac{x}{s_0}\right)^{\frac{n_1(1)-1}{2}} T^{-%
\frac{(2H+1)(n_1(1)-1)}{4}}  \notag \\
&\quad\left(1+O\left(T^{2H+1}\right)\right)  \label{E:ee11}
\end{align}
as $T\rightarrow 0$. Moreover, 
\begin{align}
z(T)^{-\frac{1}{2}}&=2\left[\frac{\lambda_1(1)T^{2H+1}+4}{%
\lambda_1(1)T^{2H+1}}\right]^{-\frac{1}{4}}  \notag \\
&=\sqrt{2}\lambda_1(1)^{\frac{1}{4}}T^{\frac{2H+1}{4}}\left(1+O%
\left(T^{2H+1}\right)\right)  \label{E:ee0}
\end{align}
and 
\begin{align}
&\exp\left\{-2z(T)\log\frac{x}{s_0}\right\}=\left(\frac{x}{s_0}\right)^{-%
\frac{\sqrt{4+\lambda_1(1)T^{2H+1}}} {2\sqrt{\lambda_1(1)} T^{H+\frac{1}{2}}}%
}  \label{E:ee2}
\end{align}
as $T\rightarrow 0$. Next, combining (\ref{E:ot2}), (\ref{E:uu4}), (\ref%
{E:ee11}), (\ref{E:ee0}), and (\ref{E:ee2}), and simplifying the resulting
expressions, we obtain formula (\ref{E:forr}).

This completes the proof of Theorem \ref{T:charc}.

\section{Asymptotic behavor of out-of-the-money call and put pricing
functions}

\label{S:call} Let $S$ be the asset price process in the model considered in
(\ref{E:1}). Define the call and the put pricing functions by 
\begin{equation*}
C(T,K)=\mathbb{E}\left[S_T-K\right]^{+}\quad\mbox{and}\quad P(T,K)=\mathbb{E}%
\left[K-S_T\right]^{+}
\end{equation*}
where $T$ is the maturity and $K$ is the strike price. Recall that for a
Gaussian stochastic volatility model with $r=0$, the asset price process $S$
is a martingale (see Lemma \ref{L:environ}). Therefore, the put/call parity
formula $C(T,K)=P(T,K)+s_0-K$ holds.

In the present section, we consider the functions $C$ and $P$ as functions
of the maturity for a fixed strike price, and we suppress the strike price
in the symbols. Our goal is to characterize the asymptotic behavior as $%
T\rightarrow 0$ of the function $T\mapsto C(T)$ for $K> s_0$ (out-of-the
money call) and of the function $T\mapsto P(T)$ for $0< K< s_0$
(out-of-the-money put).

We will first consider the call pricing function $T\mapsto C(T)$ with $K>
s_0 $. It is known that 
\begin{equation}
C(T)=\int_K^{\infty}(x-K)D_T(x)dx.  \label{E:callr}
\end{equation}
Therefore, we can use the uniform estimate in formula (\ref{E:forr}) to
characterize the small-time behavior of the call pricing function. Let us
consider the following integrals: 
\begin{align}
&I_1(T)=\int_K^{\infty}(x-K)x^{-\frac{3}{2}}\left(\log\frac{x}{s_0}\right)^{%
\frac{n_1(1)-2}{2}} \exp\left\{-2z(T)\log\frac{x}{s_0}\right\}dx  \notag \\
&=s_0^{-\frac{1}{2}}\int_K^{\infty}\left(\log\frac{x}{s_0}\right)^{\frac{%
n_1(1)-2}{2}} \exp\left\{-\left(\frac{1}{2}+2z(T)\right)\log\frac{x}{s_0}%
\right\}dx  \notag \\
&\quad-s_0^{-\frac{3}{2}}K\int_K^{\infty}\left(\log\frac{x}{s_0}\right)^{%
\frac{n_1(1)-2}{2}} \exp\left\{-\left(\frac{3}{2}+2z(T)\right)\log\frac{x}{%
s_0}\right\}dx  \label{E:noo1}
\end{align}
and 
\begin{align}
&I_2(T)=\int_K^{\infty}(x-K)x^{-\frac{3}{2}}\left(\log\frac{x}{s_0}\right)^{%
\frac{n_1(1)-3}{2}} \exp\left\{-2z(T)\log\frac{x}{s_0}\right\}dx  \notag \\
&=s_0^{-\frac{1}{2}}\int_K^{\infty}\left(\log\frac{x}{s_0}\right)^{\frac{%
n_1(1)-3}{2}} \exp\left\{-\left(\frac{1}{2}+2z(T)\right)\log\frac{x}{s_0}%
\right\}dx  \notag \\
&\quad-s_0^{-\frac{3}{2}}K\int_K^{\infty}\left(\log\frac{x}{s_0}\right)^{%
\frac{n_1(1)-3}{2}} \exp\left\{-\left(\frac{3}{2}+2z(T)\right)\log\frac{x}{%
s_0}\right\}dx,  \label{E:noo2}
\end{align}
where we use the notation in (\ref{E:red}) for the sake of shortness.

We will next make a substitution $u=(2z(T)-\frac{1}{2})\log\frac{x}{s_0}$ in
the integral on the second line in (\ref{E:noo1}). The resulting expression
is as follows: 
\begin{equation*}
s_0^{\frac{1}{2}}\left(2z(T)-\frac{1}{2}\right)^{-\frac{n_1(1)}{2}%
}\int_{\left(2z(T)-\frac{1}{2}\right)\log\frac{K}{s_0}} ^{\infty}u^{\frac{%
n_1(1)-2}{2}}e^{-u}du,
\end{equation*}
which is equal to 
\begin{equation*}
s_0^{\frac{1}{2}}\left(2z(T)-\frac{1}{2}\right)^{-\frac{n_1(1)}{2}%
}\Gamma\left(\frac{n_1(1)}{2}, \left(2z(T)-\frac{1}{2}\right)\log\frac{K}{s_0%
}\right),
\end{equation*}
where the symbol $\Gamma$ stands for the upper incomplete gamma function
defined by 
\begin{equation*}
\Gamma(s,x)=\int_x^{\infty}v^{s-1}e^{-v}dv.
\end{equation*}
Making similar transformations in the other integrals in (\ref{E:noo1}) and (%
\ref{E:noo2}), we finally obtain 
\begin{align*}
&I_1(T)=s_0^{\frac{1}{2}}\left(2z(T)-\frac{1}{2}\right)^{-\frac{n_1(1)}{2}%
}\Gamma\left(\frac{n_1(1)}{2}, \left(2z(T)-\frac{1}{2}\right)\log\frac{K}{s_0%
}\right) \\
&\quad-s_0^{-\frac{1}{2}}K\left(2z(T)+\frac{1}{2}\right)^{-\frac{n_1(1)}{2}%
}\Gamma\left(\frac{n_1(1)}{2}, \left(2z(T)+\frac{1}{2}\right)\log\frac{K}{s_0%
}\right)
\end{align*}
and 
\begin{align*}
&I_2(T)=s_0^{\frac{1}{2}}\left(2z(T)-\frac{1}{2}\right)^{-\frac{n_1(1)-1}{2}%
}\Gamma\left(\frac{n_1(1)-1}{2}, \left(2z(T)-\frac{1}{2}\right)\log\frac{K}{%
s_0}\right) \\
&\quad-s_0^{-\frac{1}{2}}K\left(2z(T)+\frac{1}{2}\right)^{-\frac{n_1(1)-1}{2}%
}\Gamma\left(\frac{n_1(1)-1}{2}, \left(2z(T)+\frac{1}{2}\right)\log\frac{K}{%
s_0}\right).
\end{align*}

It is known that 
\begin{equation}
\Gamma (s,x)=x^{s-1}e^{-x}\left( 1+(s-1)x^{-1}+O\left( x^{-2}\right) \right)
\label{E:neo1}
\end{equation}%
as $x\rightarrow \infty $. Formula (\ref{E:neo1}) can be easily derived from
the recurrence relation 
\begin{equation*}
\Gamma (s,x)=(s-1)\Gamma (s-1,x)+x^{s-1}e^{-x}
\end{equation*}%
for the upper incomplete gamma function. It follows that 
\begin{align*}
& I_{1}(T)=s_{0}^{2z(T)}K^{-2z(T)+\frac{1}{2}}\left( \log \frac{K}{s_{0}}%
\right) ^{\frac{n_{1}(1)-2}{2}} \\
& \quad \lbrack \frac{1}{2z(T)-\frac{1}{2}}\left( 1+\frac{n_{1}(1)-2}{%
2(2z(T)-\frac{1}{2})\log \frac{K}{s_{0}}}+O(T^{2H+1})\right) \\
& \quad -\frac{1}{2z(T)+\frac{1}{2}}\left( 1+\frac{n_{1}(1)-2}{2(2z(T)+\frac{%
1}{2})\log \frac{K}{s_{0}}}+O(T^{2H+1})\right) ] \\
& =s_{0}^{2z(T)}K^{-2z(T)+\frac{1}{2}}\left( \log \frac{K}{s_{0}}\right) ^{%
\frac{n_{1}(1)-2}{2}}\left( \frac{1}{4z(T)^{2}-\frac{1}{4}}+O\left( T^{3H+%
\frac{3}{2}}\right) \right)
\end{align*}%
as $T\rightarrow 0$. Therefore, 
\begin{align}
& I_{1}(T)=s_{0}^{2z(T)}K^{-2z(T)+\frac{1}{2}}\left( \log \frac{K}{s_{0}}%
\right) ^{\frac{n_{1}(1)-2}{2}}\left( 4z(T)^{2}-\frac{1}{4}\right) ^{-1} 
\notag \\
& \quad \left( 1+O\left( T^{H+\frac{1}{2}}\right) \right)  \label{E:ff1}
\end{align}%
as $T\rightarrow 0$. Similarly, 
\begin{align}
& I_{2}(T)=s_{0}^{2z(T)}K^{-2z(T)+\frac{1}{2}}\left( \log \frac{K}{s_{0}}%
\right) ^{\frac{n_{1}(1)-3}{2}}\left( 4z(T)^{2}-\frac{1}{4}\right) ^{-1} 
\notag \\
& \quad \left( 1+O\left( T^{H+\frac{1}{2}}\right) \right)  \label{E:ff2}
\end{align}%
as $T\rightarrow 0$. It is not hard to see that 
\begin{equation*}
\left( 4z(T)^{2}-\frac{1}{4}\right) ^{-1}=\lambda _{1}(1)T^{2H+1}.
\end{equation*}%
It follows from (\ref{E:ff1}) and (\ref{E:ff2}) that 
\begin{align}
& I_{1}(T)=\lambda _{1}(1)K^{\frac{1}{2}}\left( \log \frac{K}{s_{0}}\right)
^{\frac{n_{1}(1)-2}{2}}\left( \frac{s_{0}}{K}\right) ^{2z(T)}T^{2H+1}  \notag
\\
& \quad \left( 1+O\left( T^{H+\frac{1}{2}}\right) \right)  \label{E:ff3}
\end{align}%
as $T\rightarrow 0$. Similarly, 
\begin{align}
& I_{2}(T)=\lambda _{1}(1)K^{\frac{1}{2}}\left( \log \frac{K}{s_{0}}\right)
^{\frac{n_{1}(1)-3}{2}}\left( \frac{s_{0}}{K}\right) ^{2z(T)}T^{2H+1}  \notag
\\
& \quad \left( 1+O\left( T^{H+\frac{1}{2}}\right) \right)  \label{E:ff4}
\end{align}%
as $T\rightarrow 0$.

The next assertion characterizes the small-time asymptotic behavior of the
call pricing function.

\begin{theorem}
\label{T:call} Let $K> s_0$. Then the following asymptotic formula holds for
the call pricing function in the model described by (\ref{E:1}): 
\begin{align}
C(T)&=MT^{\frac{(2H+1)(4-n_1(1))}{4}}\left(\frac{s_0}{K}\right)^{%
\lambda_1(1)^{-\frac{1}{2}} T^{-H-\frac{1}{2}}}  \notag \\
&\quad\left(1+O\left(T^{\frac{2H+1}{4}}\right)\right)  \label{E:fff1}
\end{align}
as $T\rightarrow 0$, where 
\begin{align}
M&=\frac{(s_0K)^{\frac{1}{2}}}{2^{\frac{n_1(1)}{2}}\Gamma\left(\frac{n_1(1)}{%
2}\right)} \lambda_1(1)^{\frac{4-n_1(1)}{4}}\left(\log\frac{K}{s_0}\right) ^{%
\frac{n_1(1)-2}{2}}  \notag \\
&\quad\times\prod_{k=2}^{\infty}\left(\frac{\lambda_1(1)}{%
\lambda_1(1)-\rho_k(1)}\right) ^{\frac{n_k}{2}}.  \label{E:M}
\end{align}
\end{theorem}

Proof. Using (\ref{E:forr}), (\ref{E:callr}), (\ref{E:noo1}) and (\ref%
{E:noo2}), we see that 
\begin{align*}
&C(T)=\frac{\sqrt{s_0}}{2^{\frac{n_1(1)}{2}}\Gamma\left(\frac{n_1(1)}{2}%
\right)}\lambda_1(1)^{-\frac{n_1(1)}{4}} \prod_{k=2}^{\infty}\left(\frac{%
\lambda_1(1)}{\lambda_1(1)-\rho_k(1)}\right) ^{\frac{n_k}{2}} \\
&\quad T^{-\frac{(2H+1)n_1(1)}{4}} \left(1+O\left(T^{2H+1}\right)\right)%
\left[I_1(T)+O\left(T^{\frac{2H+1}{4}}I_2(T)\right)\right]
\end{align*}
as $T\rightarrow 0$. Next, (\ref{E:ff3}) and (\ref{E:ff4}), imply 
\begin{align}
&C(T)=\frac{(s_0K)^{\frac{1}{2}}}{2^{\frac{n_1(1)}{2}}\Gamma\left(\frac{%
n_1(1)}{2}\right)} \lambda_1(1)^{-\frac{n_1(1)-4}{4}}\prod_{k=2}^{\infty}%
\left(\frac{\lambda_1(1)}{\lambda_1(1)-\rho_k(1)}\right) ^{\frac{n_k}{2}} 
\notag \\
&\quad \left(\log\frac{K}{s_0}\right) ^{\frac{n_1(1)-2}{2}}T^{\frac{%
(2H+1)(4-n_1(1))}{4}}\left(\frac{s_0}{K}\right)^{2z(T)} \left(1+O\left(T^{%
\frac{2H+1}{4}}\right)\right)  \label{E:fiinn}
\end{align}
as $T\rightarrow 0$. We also have 
\begin{equation}
\sqrt{\frac{\lambda_1(1)T^{2H+1}+4}{\lambda_1(1)T^{2H+1}}}-\sqrt{\frac{4}{%
\lambda_1(1)T^{2H+1}}} =O\left(T^{H+\frac{1}{2}}\right)  \label{E:fin1}
\end{equation}
as $T\rightarrow 0$. Therefore, 
\begin{align*}
&\left(\frac{s_0}{K}\right)^{2z(T)}=\exp\left\{-2z(T)\log\frac{K}{s_0}%
\right\} \\
&=\exp\left\{-\frac{1}{2}\sqrt{\frac{\lambda_1(1)T^{2H+1}+4}{%
\lambda_1(1)T^{2H+1}}}\log\frac{K}{s_0}\right\} =\exp\left\{-\frac{1}{2}%
\sqrt{\frac{4}{\lambda_1(1)T^{2H+1}}}\log\frac{K}{s_0}\right\} \\
&\exp\left\{-\frac{1}{2}\left[\sqrt{\frac{\lambda_1(1)T^{2H+1}+4}{%
\lambda_1(1)T^{2H+1}}}-\sqrt{\frac{4}{\lambda_1(1)T^{2H+1}}}\right]\log\frac{%
K}{s_0}\right\}
\end{align*}
as $T\rightarrow 0$. Using (\ref{E:fin1}), we obtain 
\begin{align}
&\left(\frac{s_0}{K}\right)^{2z(T)}=\left(\frac{s_0}{K}\right)^{%
\lambda_1(1)^{-\frac{1}{2}}T^{-H-\frac{1}{2}}} \left(1+O\left(T^{H+\frac{1}{2%
}}\right)\right)  \label{E:ai}
\end{align}
as $T\rightarrow 0$.

Now, it is clear that Theorem \ref{T:call} follows from (\ref{E:fiinn}) and (%
\ref{E:ai}).

The next statement allows us to recover the self-similarity index $H$ from
the asymptotics of the call pricing function.

\begin{corollary}
\label{C:corrola} Under the conditions in Theorem \ref{T:call}, for every $%
K> s_0$, 
\begin{equation}
H=\lim_{T\rightarrow 0}\frac{\log\log\frac{1}{C(T,K)}}{\log\frac{1}{T}}-%
\frac{1}{2}.  \label{E:blesk}
\end{equation}
\end{corollary}

Proof. It follows from (\ref{E:fff1}) that 
\begin{align}
\log\frac{1}{C(T)}&=\log\frac{1}{M}+\frac{(2H+1)(4-n_1(1))}{4}\log\frac{1}{T}
\notag \\
&\quad+\lambda_1(1)^{-\frac{1}{2}} T^{-H-\frac{1}{2}}\log\frac{K}{s_0}%
+O\left(T^{\frac{2H+1}{4}}\right)  \label{E:e1}
\end{align}
as $T\rightarrow 0$. Hence, 
\begin{align}
&\log\log\frac{1}{C(T)}=\log\left[\lambda_1(1)^{-\frac{1}{2}}T^{-H-\frac{1}{2%
}}\log\frac{K}{s_0}\right]  \notag \\
&+\log\left(1+O\left(T^{H+\frac{1}{2}}+T^{H+\frac{1}{2}}\log\frac{1}{T}
+T^{H+\frac{1}{2}}O\left(T^{\frac{2H+1}{4}}\right)\right)\right)  \notag \\
&=\left(H+\frac{1}{2}\right)\log\frac{1}{T}+\log\left[\lambda_1(1)^{-\frac{1%
}{2}}\log\frac{K}{s_0}\right] +O\left(T^{H+\frac{1}{2}}\log\frac{1}{T}\right)
\label{E:e2}
\end{align}
as $T\rightarrow 0$.

Now, it is clear that (\ref{E:blesk}) follows from the previous formula.

Next, we turn our attention to the out-of-the-money put pricing function $%
T\mapsto P(T)$ with $0< K< s_0$. The asymptotic behavior of the put pricing
function with $0< K< s_0$ will be characterized using the symmetry
properties of the model in (\ref{E:1}). In (\cite{G}, Lemma 9.25), several
equivalent conditions are given for the symmetry of a stochastic volatility
model. One of them is as follows (see (9.79) in \cite{G}): 
\begin{equation}
D_T(x)=\left(\frac{s_0}{x}\right)^3D_T\left(\frac{s_0^2}{x}\right)
\label{E:sy1}
\end{equation}
for all $x> 0$ and $T> 0$. It is clear that for the model described by (\ref%
{E:1}), the previous equality can be derived from formula (\ref{E:sictr}).
Next, using Theorem \ref{T:charc} and (\ref{E:sy1}), we establish the
following proposition.

\begin{theorem}
\label{T:charco} Let $0<\varepsilon< s_0$ and $0< x< s_0-\varepsilon$. Then
as $T\rightarrow 0$, the following asymptotic formula holds for the asset
price density $D_T$ in the model described by (\ref{E:svm}): 
\begin{align}
D_T(x)&=\frac{\sqrt{s_0}}{2^{\frac{n_1(1)}{2}}\Gamma\left(\frac{n_1(1)}{2}%
\right)}\lambda_1(1)^{-\frac{n_1(1)}{4}} \prod_{k=2}^{\infty}\left(\frac{%
\lambda_1(1)}{\lambda_1(1)-\rho_k(1)}\right) ^{\frac{n_k}{2}}x^{-\frac{3}{2}}
\notag \\
&\quad\times\left(\log\frac{s_0}{x}\right)^{\frac{n_1(1)-2}{2}}T^{-\frac{%
(2H+1)n_1(1)}{4}} \left(\frac{s_0}{x}\right)^{-\frac{\sqrt{%
4+\lambda_1(1)T^{2H+1}}} {2\sqrt{\lambda_1(1)}T^{H+\frac{1}{2}}}}  \notag \\
&\quad\times\left(1+O\left(T^{2H+1}\right)\right)\left(1+O_{\varepsilon}%
\left(T^{\frac{2H+1}{4}} \left(\log\frac{s_0}{x}\right)^{-\frac{1}{2}%
}\right)\right).  \label{E:forra}
\end{align}
\end{theorem}

Since the model that we are studying is symmetric, 
\begin{equation}
P(T,K)=\frac{K}{s_0} C\left(T,\frac{s_0^2}{K}\right).  \label{E:kon1}
\end{equation}
(see condition 3 in Lemma 9.25 in \cite{G}).

The next assertion follows from Theorem \ref{T:call} and (\ref{E:kon1}).

\begin{theorem}
\label{T:put} Let $0<K<s_{0}$. Then the following asymptotic formula holds
for the put pricing function in the model described by (\ref{E:1}): 
\begin{align}
P(T)& =\tilde{M}T^{\frac{(2H+1)(4-n_{1}(1))}{4}}\left( \frac{K}{s_{0}}%
\right) ^{\lambda _{1}(1)^{-\frac{1}{2}}T^{-H-\frac{1}{2}}}  \notag \\
& \quad \left( 1+O\left( T^{\frac{2H+1}{4}}\right) \right)  \label{E:fof1}
\end{align}%
as $T\rightarrow 0$, where the constant $\tilde{M}$ is given by 
\begin{align}
\tilde{M}& =\frac{(s_{0}K)^{\frac{1}{2}}}{2^{\frac{n_{1}(1)}{2}}\Gamma
\left( \frac{n_{1}(1)}{2}\right) }\lambda _{1}(1)^{\frac{4-n_{1}(1)}{4}%
}\left( \log \frac{s_{0}}{K}\right) ^{\frac{n_{1}(1)-2}{2}}  \notag \\
& \quad \times \prod_{k=2}^{\infty }\left( \frac{\lambda _{1}(1)}{\lambda
_{1}(1)-\rho _{k}(1)}\right) ^{\frac{n_{k}}{2}}.  \label{E:Mtilde}
\end{align}
\end{theorem}

Next, using the same reasoning as in the proof of Corollary \ref{C:corrola},
we obtain the following statement.

\begin{corollary}
\label{C:toyota} Under the conditions in Theorem \ref{T:put}, for every $0<
K< s_0$, 
\begin{equation}
H=\lim_{T\rightarrow 0}\frac{\log\log\frac{1}{P(T,K)}}{\log\frac{1}{T}}-%
\frac{1}{2}.  \label{E:bleski}
\end{equation}
\end{corollary}

\section{Asymptotic behavior of the implied volatility}

\label{S:gl} Theorems \ref{T:call} and \ref{T:put} characterize the
small-time behavior of the call and put pricing functions in a stochastic
volatility model with centered Gaussian self-similar volatility. In the
present section, we study the small-time behavior of the implied volatility
in such a model. We will use some of the results obtained by Gao and Lee in 
\cite{GL}. Gao and Lee establish certain asymptotic relations between the
implied volatility and the call pricing function under very general
conditions. They consider various asymptotic regimes, e.g., the extreme
strike, the small/large time, or mixed regimes. Of our interest is formula
(7.11) in Corollary 7.3 in \cite{GL}, providing an asymptotic formula
characterizing the small-time asymptotic behavior of the implied volatility
in terms of the call pricing function. It follows from this formula that if $%
K\neq s_0$, then 
\begin{equation*}
\sqrt{T}I(T,K)=\frac{\left|\log\frac{K}{s_0}\right|}{\sqrt{2\left|\log\frac{1%
}{C(T,K)}\right|}}\left(1+O\left(\frac{\log\left|\log\frac{1}{C(T,K)}\right|%
} {\left|\log\frac{1}{C(T,K)}\right|}\right)\right)
\end{equation*}
as $T\rightarrow 0$. Therefore, 
\begin{equation}
I(T,K)=\frac{\left|\log\frac{K}{s_0}\right|}{\sqrt{2T\left|\log\frac{1}{%
C(T,K)}\right|}} +O\left(\frac{\log\left|\log\frac{1}{C(T,K)}\right|} {\sqrt{%
T}\left|\log\frac{1}{C(T,K)}\right|^{\frac{3}{2}}}\right)  \label{E:ci1}
\end{equation}
as $T\rightarrow 0$.

The following assertion can be derived from (\ref{E:fff1}) and (\ref{E:ci1}).

\begin{theorem}
\label{T:fiv} Let $K> s_0$. Then the following asymptotic formula holds for
the implied volatility in the model described by (\ref{E:1}): 
\begin{equation}
I(T)=\frac{\lambda_1(1)^{\frac{1}{4}}\sqrt{\log\frac{K}{s_0}}}{\sqrt{2}}T^{%
\frac{2H-1}{4}}+O\left(T^{\frac{6H+1}{4}}\log\frac{1}{T}\right)
\label{E:fiv1}
\end{equation}
as $T\rightarrow 0$.
\end{theorem}

Proof. It follows from (\ref{E:e1}) and (\ref{E:e2}) that 
\begin{equation*}
\log\frac{1}{C(T)}\approx T^{-H-\frac{1}{2}}
\end{equation*}
and 
\begin{equation*}
\log\log\frac{1}{C(T)}\approx\log\frac{1}{T}
\end{equation*}
as $T\rightarrow 0$. Moreover, the mean value theorem implies that 
\begin{align*}
\left(\log\frac{1}{C(T)}\right)^{-\frac{1}{2}}&=\left(\lambda_1(1)^{-\frac{1%
}{2}} T^{-H-\frac{1}{2}}\log\frac{K}{s_0}\right)^{-\frac{1}{2}}+O\left(T^{%
\frac{6H+3}{4}}\log\frac{1}{T}\right)  \notag \\
&=\lambda_1(1)^{\frac{1}{4}}\left(\log\frac{K}{s_0}\right)^{-\frac{1}{2}}T^{%
\frac{2H+1}{4}} +O\left(T^{\frac{6H+3}{4}}\log\frac{1}{T}\right)
\end{align*}
as $T\rightarrow 0$. Now it is not hard to see that (\ref{E:fiv1}) follows
from (\ref{E:ci1}) and the previous formulas.

\begin{remark}
\label{R:prove} \textrm{Assume $K> s_0$. It follows from Theorem \ref{T:fiv}
that if the Hurst index satisfies $0< H<\frac{1}{2}$, then the implied
volatility $T\mapsto I(K,T)$ is singular at $T=0$, and it behaves near zero
like the function $T\mapsto T^{\frac{2H-1}{4}}$. For standard Brownian
motion, $H=\frac{1}{2}$, and we have 
\begin{equation*}
\lim_{T\rightarrow 0}I(K,T)=\frac{\lambda_1(1)^{\frac{1}{4}}\sqrt{\log\frac{K%
}{s_0}}}{\sqrt{2}}.
\end{equation*}
Finally, for $\frac{1}{2}< H< 1$, the implied volatility $T\mapsto I(K,T)$
tends to zero like the function $T\mapsto T^{\frac{2H-1}{4}}$. }
\end{remark}

The next statement is a corollary to Theorem \ref{T:fiv}. It provides a
representation of the self-similarity index in terms of the implied
volatility.

\begin{corollary}
\label{C:fiv} Let $K> s_0$. Then the following equality holds: 
\begin{equation}
H=2\lim_{T\rightarrow 0}\frac{\log\frac{1}{I(T,K)}}{\log\frac{1}{T}}+\frac{1%
}{2}.  \label{E:fiv2}
\end{equation}
\end{corollary}

In the case where $0< K< s_0$, Theorem \ref{T:fiv}, Corollary \ref{C:fiv},
and the symmetry condition 
\begin{equation*}
I(T,K)=I\left(T,\frac{s_0^2}{K}\right)
\end{equation*}
(see \cite{G}, Lemma 9.25) imply the following assertions.

\begin{theorem}
\label{T:fivy} Let $0<K< s_0$. Then the following asymptotic formula holds
for the implied volatility in the model described by (\ref{E:1}): 
\begin{equation}
I(T)=\frac{\lambda_1(1)^{\frac{1}{4}}\sqrt{\log\frac{s_0}{K}}}{\sqrt{2}}T^{%
\frac{2H-1}{4}}+O\left(T^{\frac{6H+1}{4}}\log\frac{1}{T}\right)
\label{E:fivy1}
\end{equation}
as $T\rightarrow 0$.
\end{theorem}

\begin{corollary}
\label{C:fivy} Let $0< K< s_0$. Then equality (\ref{E:fiv2}) holds for the
self-similarity index $H$.
\end{corollary}

\section{At-the-money options}

\label{S:atthe} In this section, we consider a stochastic volatlity model,
in which the volatility process $X^{(H)}$ is an adapted $H$-self-similar
Gaussian process. As before, we assume $r=0$. Let us also suppose $K=s_0$
(at-the-money case). Note that here we do not assume that the volatility
process is centered.

Using (\ref{E:sictr}) and the formula 
\begin{equation*}
C(T,K)=\int_K^{\infty}(x-K)D_T(x)dx,
\end{equation*}
we obtain the following equalities for the at-the-money call: 
\begin{align*}
&C(T,s_0)=\frac{\sqrt{s_0}}{\sqrt{2\pi}}T^{-H-\frac{1}{2}}\int_0^{\infty}
u^{-1}\exp\left\{-\frac{T^{2H+1}u^2}{8}\right\}\widetilde{p}_1(u)du \\
&\times\int_{s_0}^{\infty}(x-s_0) x^{-\frac{3}{2}}\exp\left\{-\frac{\log^2%
\frac{x}{s_0}}{2T^{2H+1}u^2}\right\}dx \\
&=\frac{\sqrt{s_0}}{\sqrt{2\pi}}T^{-H-\frac{1}{2}}\int_0^{\infty}
u^{-1}\exp\left\{-\frac{T^{2H+1}u^2}{8}\right\}\widetilde{p}_1(u)du \\
&\times\left[\int_{s_0}^{\infty} x^{-\frac{1}{2}}\exp\left\{-\frac{\log^2%
\frac{x}{s_0}}{2T^{2H+1}u^2}\right\}dx -s_0\int_{s_0}^{\infty} x^{-\frac{3}{2%
}}\exp\left\{-\frac{\log^2\frac{x}{s_0}}{2T^{2H+1}u^2}\right\}dx\right].
\end{align*}
It follows from the previous formula that 
\begin{align}
C(T,s_0)&=\frac{s_0}{\sqrt{2\pi}}T^{-H-\frac{1}{2}}\int_0^{\infty}
u^{-1}\exp\left\{-\frac{T^{2H+1}u^2}{8}\right\}\widetilde{p}_1(u)  \notag \\
&\times\left[\Phi_1(T,u)-\Phi_2(T,u)\right]du,  \label{E:sict}
\end{align}
where 
\begin{equation}
\Phi_1(T,u)=\int_1^{\infty} y^{-\frac{1}{2}}\exp\left\{-\frac{\log^2y}{%
2T^{2H+1}u^2}\right\}dy  \label{E:sict1}
\end{equation}
and 
\begin{equation}
\Phi_2(T,u)=\int_1^{\infty} y^{-\frac{3}{2}}\exp\left\{-\frac{\log^2y}{%
2T^{2H+1}u^2}\right\}dy.  \label{E:sict2}
\end{equation}

Our next goal is to estimate the functions $\Phi _{1}$ and $\Phi _{2}$
defined in (\ref{E:sict1}) and (\ref{E:sict2}). We have 
\begin{align*}
& \Phi _{1}(T,u)=\int_{0}^{\infty }\exp \left\{ -\left[ \frac{w^{2}}{%
2T^{2H+1}u^{2}}-\frac{w}{2}\right] \right\} dw \\
& =\exp \left\{ \frac{T^{2H+1}u^{2}}{8}\right\} \int_{0}^{\infty }\exp
\left\{ -\frac{1}{2T^{2H+1}u^{2}}\left( w-\frac{T^{2H+1}u^{2}}{2}\right)
^{2}\right\} dw \\
& =\exp \left\{ \frac{T^{2H+1}u^{2}}{8}\right\} \int_{-\frac{1}{2}%
T^{2H+1}u^{2}}^{\infty }\exp \left\{ -\frac{1}{2T^{2H+1}u^{2}}z^{2}\right\}
dz \\
& =T^{H+\frac{1}{2}}u\exp \left\{ \frac{T^{2H+1}u^{2}}{8}\right\} \int_{-%
\frac{1}{2}T^{H+\frac{1}{2}}u}^{\infty }\exp \left\{ -\frac{y^{2}}{2}%
\right\} dy.
\end{align*}%
Similarly, 
\begin{equation*}
\Phi _{2}(T,u)=T^{H+\frac{1}{2}}u\exp \left\{ \frac{T^{2H+1}u^{2}}{8}%
\right\} \int_{\frac{1}{2}T^{H+\frac{1}{2}}u}^{\infty }\exp \left\{ -\frac{%
y^{2}}{2}\right\} dy.
\end{equation*}%
Therefore 
\begin{equation}
\Phi _{1}(T,u)-\Phi _{2}(T,u)=2T^{H+\frac{1}{2}}u\exp \left\{ \frac{%
T^{2H+1}u^{2}}{8}\right\} \int_{0}^{\frac{1}{2}T^{H+\frac{1}{2}}u}\exp
\left\{ -\frac{y^{2}}{2}\right\} dy.  \label{E:aa}
\end{equation}

The next lemma will be useful in the sequel. It will allow us to estimate
the integral in (\ref{E:aa}).

\begin{lemma}
\label{L:a} Let $0< a< 1$. Then the following inequalities are valid: 
\begin{equation}
a-\frac{a^3}{6}\le\int_0^a\exp\left\{-\frac{y^2}{2}\right\}dy\le a-\frac{a^3%
}{6}+\frac{a^5}{40}.  \label{E:a1}
\end{equation}
On the other hand, if $a\ge 1$, then 
\begin{align}
\frac{\sqrt{\pi}}{\sqrt{2}}-\frac{1}{a}\exp\left\{-\frac{a^2}{2}%
\right\}&\le\int_0^a\exp\left\{-\frac{y^2}{2}\right\}dy  \notag \\
&\le\frac{\sqrt{\pi}}{\sqrt{2}}-\frac{a}{a^2+1}\exp\left\{-\frac{a^2}{2}%
\right\}.  \label{E:a2}
\end{align}
\end{lemma}

Proof. The inequalities in (\ref{E:a1}) can be established using the Taylor
expansion with two and three terms.

To prove the estimates in (\ref{E:a2}), we use the following known
inequalities: 
\begin{equation}
\frac{x}{x^{2}+1}\exp \left\{ -\frac{x^{2}}{2}\right\} \leq \int_{x}^{\infty
}\exp \left\{ -\frac{y^{2}}{2}\right\} dy\leq \frac{1}{x}\exp \left\{ -\frac{%
x^{2}}{2}\right\} ,  \label{E:a3}
\end{equation}%
for all $x>0$. The previous inequalities follow from stronger estimates
formulated in \cite{AS}, 7.1.13. Now, (\ref{E:a2}) can be derived from (\ref%
{E:a3}) and the equality 
\begin{equation*}
\int_{0}^{a}\exp \left\{ -\frac{y^{2}}{2}\right\} dy=\frac{\sqrt{\pi }}{%
\sqrt{2}}-\int_{a}^{\infty }\exp \left\{ -\frac{y^{2}}{2}\right\} dy.
\end{equation*}

This completes the proof of Lemma \ref{L:a}.

The next assertion provides estimates for the at-the-money call.

\begin{theorem}
\label{T:b} The following inequalities are true for every $T> 0$: 
\begin{equation*}
U_1(T)\le C(T,s_0)\le U_2(T),
\end{equation*}
where 
\begin{align*}
&U_1(T)=\frac{s_0}{\sqrt{2\pi}}T^{H+\frac{1}{2}}\int_0^{\infty}\widetilde{p}%
_1(u)udu -\frac{s_0}{24\sqrt{2\pi}}T^{3H+\frac{3}{2}}\int_0^{\infty}%
\widetilde{p}_1(u)u^3du \\
&+\frac{2s_0}{\sqrt{2\pi}T^{H+\frac{1}{2}}}\int_2^{\infty}\widetilde{p}%
_1\left(\frac{v}{T^{H+\frac{1}{2}}}\right) \left[\frac{\sqrt{\pi}}{\sqrt{2}}-%
\frac{v}{2}+\frac{v^3}{48}-\frac{2}{v}\exp\left\{-\frac{v^2}{8}\right\}%
\right]dv
\end{align*}
and 
\begin{align*}
&U_2(T)=\frac{s_0}{\sqrt{2\pi}}T^{H+\frac{1}{2}}\int_0^{\infty}\widetilde{p}%
_1(u)udu -\frac{s_0}{24\sqrt{2\pi}}T^{3H+\frac{3}{2}}\int_0^{\infty}%
\widetilde{p}_1(u)u^3du \\
&+\frac{s_0}{640\sqrt{2\pi}}T^{5H+\frac{5}{2}}\int_0^{\infty}\widetilde{p}%
_1(u)u^5du +\frac{2s_0}{\sqrt{2\pi}T^{H+\frac{1}{2}}}\int_2^{\infty}%
\widetilde{p}_1\left(\frac{v}{T^{H+\frac{1}{2}}}\right) \\
&\left[\frac{\sqrt{\pi}}{\sqrt{2}}-\frac{v}{2}+\frac{v^3}{48}-\frac{v^5}{1280%
} -\frac{2v}{v^2+4}\exp\left\{-\frac{v^2}{8}\right\}\right]dv
\end{align*}
\end{theorem}

Proof. It follows from (\ref{E:sict}), (\ref{E:aa}) and Lemma \ref{L:a} that 
\begin{align}
&C(T,s_0)\le\frac{s_0}{\sqrt{2\pi}} \int_0^{\frac{2}{T^{H+\frac{1}{2}}}}%
\widetilde{p}_1(u)  \notag \\
&\left[T^{H+\frac{1}{2}}u-\frac{1}{24}T^{3H+\frac{3}{2}}u^3+\frac{1}{640}%
T^{5H+\frac{5}{2}}u^5\right]du  \notag \\
&+\frac{2s_0}{\sqrt{2\pi}}\int_{\frac{2}{T^{H+\frac{1}{2}}}}^{\infty}%
\widetilde{p}_1(u) \left[\frac{\sqrt{\pi}}{\sqrt{2}}-\frac{2T^{H+\frac{1}{2}%
}u}{T^{2H+1}u^2+4}\exp\left\{-\frac{T^{2H+1}u^2}{8}\right\}\right]du  \notag
\\
&=\frac{s_0}{\sqrt{2\pi}}\int_0^{\infty}\widetilde{p}_1(u) \left[T^{H+\frac{1%
}{2}}u-\frac{1}{24}T^{3H+\frac{3}{2}}u^3+\frac{1}{640}T^{5H+\frac{5}{2}}u^5%
\right]du  \notag \\
&-\frac{2s_0}{\sqrt{2\pi}}\int_{\frac{2}{T^{H+\frac{1}{2}}}}^{\infty}%
\widetilde{p}_1(u) \left[\frac{1}{2}T^{H+\frac{1}{2}}u-\frac{1}{48}T^{3H+%
\frac{3}{2}}u^3+\frac{1}{1280}T^{5H+\frac{5}{2}}u^5\right]du  \notag \\
&+\frac{2s_0}{\sqrt{2\pi}}\int_{\frac{2}{T^{H+\frac{1}{2}}}}^{\infty}%
\widetilde{p}_1(u) \left[\frac{\sqrt{\pi}}{\sqrt{2}}-\frac{2T^{H+\frac{1}{2}%
}u}{T^{2H+1}u^2+4}\exp\left\{-\frac{T^{2H+1}u^2}{8}\right\}\right]du.
\label{E:bb}
\end{align}
and 
\begin{align}
&C(T,s_0)\ge\frac{s_0}{\sqrt{2\pi}}\int_0^{\frac{2}{T^{H+\frac{1}{2}}}}%
\widetilde{p}_1(u) \left[T^{H+\frac{1}{2}}u-\frac{1}{24}T^{3H+\frac{3}{2}}u^3%
\right]du  \notag \\
&+\frac{2s_0}{\sqrt{2\pi}}\int_{\frac{2}{T^{H+\frac{1}{2}}}}^{\infty}%
\widetilde{p}_1(u) \left[\frac{\sqrt{\pi}}{\sqrt{2}}-\frac{2}{T^{H+\frac{1}{2%
}}u}\exp\left\{-\frac{T^{2H+1}u^2}{8}\right\}\right]du  \notag \\
&=\frac{s_0}{\sqrt{2\pi}}\int_0^{\infty}\widetilde{p}_1(u) \left[T^{H+\frac{1%
}{2}}u-\frac{1}{24}T^{3H+\frac{3}{2}}u^3\right]du  \notag \\
&-\frac{2s_0}{\sqrt{2\pi}}\int_{\frac{2}{T^{H+\frac{1}{2}}}}^{\infty}%
\widetilde{p}_1(u) \left[\frac{1}{2}T^{H+\frac{1}{2}}u-\frac{1}{48}T^{3H+%
\frac{3}{2}}u^3\right]du  \notag \\
&+\frac{2s_0}{\sqrt{2\pi}}\int_{\frac{2}{T^{H+\frac{1}{2}}}}^{\infty}%
\widetilde{p}_1(u) \left[\frac{\sqrt{\pi}}{\sqrt{2}}-\frac{2}{T^{H+\frac{1}{2%
}}u}\exp\left\{-\frac{T^{2H+1}u^2}{8}\right\}\right]du.  \label{E:bbb}
\end{align}

Now, it is not hard to see, making the substitution $v=T^{H+\frac{1}{2}}u$,
that Theorem \ref{T:b} follows from (\ref{E:bb}) and (\ref{E:bbb}).

The next statement characterizes the small-time asymptotic behavior of the
at-the-money call pricing function in a Gaussian self-similar stochastic
volatility model.

\begin{corollary}
\label{C:bis} The following formula holds as $T\rightarrow 0$: 
\begin{align}
&C(T,s_0)=c_1T^{H+\frac{1}{2}} -c_2T^{3H+\frac{3}{2}}+O\left(T^{5H+\frac{5}{2%
}}\right),  \label{E:erore}
\end{align}
where 
\begin{equation}
c_1=\frac{s_0}{\sqrt{2\pi}}\int_0^{\infty}p_1(u)u^{\frac{1}{2}}du
\label{E:c1}
\end{equation}
and 
\begin{equation}
c_2=\frac{s_0}{24\sqrt{2\pi}}\int_0^{\infty}p_1(u)u^{\frac{3}{2}}du.
\label{E:c2}
\end{equation}
\end{corollary}

Proof. For a centered volatility process $X$, we will use formula (\ref%
{E:ot1})). In the case of a noncentered volatility process $X$, we need the
following formula: 
\begin{align}
\widetilde{p}_{1}(x)& =2Cx^{\frac{n_{1}(1)-1}{2}}\exp \left\{ \sqrt{\frac{%
\delta (1)}{\lambda _{1}(1)}}x\right\} \exp \left\{ -\frac{x^{2}}{2\lambda
_{1}(1)}\right\}  \notag \\
& \quad \times \left( 1+O\left( x^{-1}\right) \right)  \label{E:finneg}
\end{align}%
as $x\rightarrow \infty $, where the constant $C$ is given by (\ref{E:rro}).
Formula (\ref{E:finneg}) now derives easily from (\ref{E:h}) and (\ref%
{E:finny}).

It follows from Theorem \ref{T:b} that 
\begin{align}
&C(T,s_0)-U_1(T)\le U_2(T)-U_1(T)  \notag \\
&\le\frac{s_0}{640\sqrt{2\pi}} T^{5H+\frac{5}{2}}\int_0^{\infty}\widetilde{p}%
_1(u)u^5du  \notag \\
&+\frac{2s_0}{\sqrt{2\pi}T^{H+\frac{1}{2}}}\int_2^{\infty}\widetilde{p}%
_1\left(\frac{v}{T^{H+\frac{1}{2}}}\right)  \notag \\
&\left[\frac{2}{v}\exp\left\{-\frac{v^2}{8}\right\} +\frac{2v}{v^2+4}%
\exp\left\{-\frac{v^2}{8}\right\}+\frac{v^5}{1280}\right]dv.  \label{E:nak}
\end{align}

Let us next suppose the process $X$ is centered. Then, using (\ref{E:ot1}),
we see that for $v> 2$ and for sufficiently small values of $T$, 
\begin{align}
\frac{1}{T^{H+\frac{1}{2}}}\widetilde{p}_1\left(\frac{v}{T^{H+\frac{1}{2}}}%
\right) &\le\alpha \left(\frac{v}{T^{H+\frac{1}{2}}}\right)^{n_1(1)-1}\frac{1%
}{T^{H+\frac{1}{2}}} \exp\left\{-\frac{v^2}{2\lambda_1(1)T^{2H+1}}\right\} 
\notag \\
&\le\alpha\frac{1}{T^{H+\frac{1}{2}}}\exp\left\{-\frac{v^2}{%
4\lambda_1(1)T^{2H+1}}\right\}  \notag \\
&\le\alpha\frac{1}{T^{H+\frac{1}{2}}}\exp\left\{-\frac{1}{%
2\lambda_1(1)T^{2H+1}}\right\} \exp\left\{-\frac{v^2}{8\lambda_1(1)}\right\}
\notag \\
&\le\alpha\exp\left\{-\frac{1}{4\lambda_1(1)T^{2H+1}}\right\} \exp\left\{-%
\frac{v^2}{8\lambda_1(1)}\right\}.  \label{E:nak0}
\end{align}
Here $\alpha> 0$ is a constant that may change from line to line.

Now assume the process $X$ is noncentered. Then for $v>2$ and for
sufficiently small $T$, 
\begin{align}
\frac{1}{T^{H+\frac{1}{2}}}\widetilde{p}_{1}\left( \frac{v}{T^{H+\frac{1}{2}}%
}\right) & \leq \alpha \left( \frac{v}{T^{H+\frac{1}{2}}}\right) ^{\frac{%
n_{1}(1)-1}{2}}\frac{1}{T^{H+\frac{1}{2}}}\exp \left\{ \sqrt{\frac{\delta (1)%
}{\lambda _{1}(1)}}\frac{v}{T^{H+\frac{1}{2}}}\right\}  \notag \\
& \exp \left\{ -\frac{v^{2}}{2\lambda _{1}(1)T^{2H+1}}\right\}  \notag \\
& \leq \alpha \frac{1}{T^{H+\frac{1}{2}}}\exp \left\{ -\frac{v^{2}}{4\lambda
_{1}(1)T^{2H+1}}\right\}  \notag \\
& \leq \alpha \exp \left\{ -\frac{1}{4\lambda _{1}(1)T^{2H+1}}\right\} \exp
\left\{ -\frac{v^{2}}{8\lambda _{1}(1)}\right\} .  \label{E:nakl0}
\end{align}

Finally, taking into account (\ref{E:nak}), (\ref{E:nak0}), and (\ref%
{E:nakl0}), we obtain 
\begin{equation}
C(T,s_0)-U_1(T)=O\left(T^{5H+\frac{5}{2}}\right)  \label{E:nak1}
\end{equation}
as $T\rightarrow 0$. Now, it is not hard to see, using the definition of $%
U_1 $, (\ref{E:nak0}), and (\ref{E:nak1}) that 
\begin{equation*}
C(T,s_0)=b_1T^{H+\frac{1}{2}} -b_2T^{3H+\frac{3}{2}}+O\left(T^{5H+\frac{5}{2}%
}\right),
\end{equation*}
where 
\begin{equation*}
b_1=\frac{s_0}{\sqrt{2\pi}}\int_0^{\infty}\widetilde{p}_1(u)udu
\end{equation*}
and 
\begin{equation*}
b_2=\frac{s_0}{24\sqrt{2\pi}}\int_0^{\infty}\widetilde{p}_1(u)u^3du.
\end{equation*}
Finally, using the equality $\widetilde{p}_1(u)=2up_1(u^2)$, we obtain $%
b_i=c_i$ for $i=1,2$.

This completes the proof of Corollary \ref{C:bis}.

\section{Implied volatility in at-the-money regime}

\label{S:atmiv} The Black-Scholes call pricing function for $r=0$ and $K=s_0$
is given by 
\begin{equation*}
C_{BS}(T,s_0,\sigma)=\frac{s_0}{\sqrt{2\pi}}\int_{-\frac{\sigma\sqrt{T}}{2}%
}^{\frac{\sigma\sqrt{T}}{2}} e^{-\frac{y^2}{2}}dy=s_0\frac{2}{\sqrt{\pi}}%
\int_0^{\frac{\sigma\sqrt{T}}{2\sqrt{2}}}e^{-x^2}dx.
\end{equation*}
Hence, 
\begin{equation}
C_{BS}(T,s_0,\sigma)=s_0\,\mbox{erf}\left(\frac{\sigma\sqrt{T}}{2\sqrt{2}}%
\right),  \label{E:erf}
\end{equation}
where erf is the error function defined by $\mbox{erf}(u)=\frac{2}{\sqrt{\pi}%
}\int_0^ue^{-x^2}dx$. The error function is a strictly increasing continuous
function from $[0,\infty)$ onto $[0,1)$. Its inverse function is denoted by $%
\mbox{erf}^{-1}$. It is known that the inverse error function has the
following Maclorin's expansion: 
\begin{equation}
\mbox{erf}^{-1}(z)=\frac{\sqrt{\pi}}{2}\left(z+\frac{\pi}{12}z^3+\frac{7\pi^2%
}{480}z^5+\cdots\right),\quad 0\le z\le 1  \label{E:mcl}
\end{equation}
(see \cite{}). It follows from the definition of the implied volatility that 
\begin{equation*}
C_{BS}(T,s_0,I(T,s_0))=C(T,s_0).
\end{equation*}
Therefore, (\ref{E:erf}) implies 
\begin{equation*}
I(T,s_0)=\frac{2\sqrt{2}}{\sqrt{T}}\mbox{erf}^{-1}\left(\frac{C(T,s_0)}{s_0}%
\right).
\end{equation*}
Next, using (\ref{E:mcl}), we obtain 
\begin{equation}
I(T,s_0)=\frac{\sqrt{2\pi}}{\sqrt{T}}\left[\frac{C(T,s_0)}{s_0}+\frac{\pi}{12%
}\frac{C(T,s_0)^3}{s_0^3}+ O\left(C(T,s_0)^5\right)\right]  \label{E:err}
\end{equation}
as $T\rightarrow 0$.

Now, we are ready to characterize the small-time asymptotic behavior of the
implied volatility in at-the-money regime.

\begin{theorem}
\label{T:aiv} The following asymptotic formula holds as $T\rightarrow 0$: 
\begin{align}
I(T,s_0)&=T^H\int_0^{\infty}p_1(u)u^{\frac{1}{2}}du  \notag \\
&\quad+T^{3H+1}\frac{1}{24}\left[\left( \int_0^{\infty}p_1(u)u^{\frac{1}{2}%
}du\right)^3-\int_0^{\infty}p_1(u)u^{\frac{3}{2}}du\right]  \notag \\
&\quad+O\left(T^{5H+2}\right).  \label{E:ft}
\end{align}
\end{theorem}

Proof. Our first goal is to obtain an asymptotic formula for the implied
volatility with error term of the order $O\left(T^{5H+2}\right)$, by using
formula (\ref{E:erore}) in (\ref{E:err}). Following this plan, we obtain 
\begin{align}
&I(T,s_0)=\frac{\sqrt{2\pi}}{s_0\sqrt{T}}\left(c_1T^{H+\frac{1}{2}}
-c_2T^{3H+\frac{3}{2}}+O\left(T^{5H+\frac{5}{2}}\right)\right)  \notag \\
&+\frac{\pi\sqrt{2\pi}}{12s_0^3\sqrt{T}}\left(c_1T^{H+\frac{1}{2}} -c_2T^{3H+%
\frac{3}{2}}+O\left(T^{5H+\frac{5}{2}}\right)\right)^3+O\left(T^{5H+2}\right)
\notag \\
&=\frac{\sqrt{2\pi}c_1}{s_0}T^H+\left(\frac{\pi\sqrt{2\pi}c_1^3}{12s_0^3} -%
\frac{\sqrt{2\pi}c_2}{s_0}\right)T^{3H+1} +O\left(T^{5H+2}\right)
\label{E:finni}
\end{align}
as $T\rightarrow 0$. Now, it is not difficult to see that formula (\ref{E:ft}%
) follows from (\ref{E:c1}), (\ref{E:c2}), and (\ref{E:finni}).

This completes the proof of Theorem \ref{T:aiv}.

\begin{remark}
\label{R:intv} \textrm{It is clear that the following formulas are valid for
the integrals in (\ref{E:ft}): 
\begin{equation*}
\mu _{1/2}:=\int_{0}^{\infty }p_{1}(u)u^{\frac{1}{2}}du=\mathbb{E}\left[
\left( \int_{0}^{1}X_{s}^{2}ds\right) ^{\frac{1}{2}}\right]
\end{equation*}%
and 
\begin{equation*}
\mu _{3/2}:=\int_{0}^{\infty }p_{1}(u)u^{\frac{3}{2}}du=\mathbb{E}\left[
\left( \int_{0}^{1}X_{s}^{2}ds\right) ^{\frac{3}{2}}\right] .
\end{equation*}%
}
\end{remark}

Theorem \ref{T:aiv} allows us to recover the self-similarity index $H$
knowing the small-time behavior of the at-the-money implied volatility.

\begin{theorem}
\label{T:recover} The following formula holds: 
\begin{equation*}
H=\lim_{T\rightarrow 0}\frac{\log \frac{1}{I(T,s_{0})}}{\log \frac{1}{T}}.
\end{equation*}
\end{theorem}

\section{Numerical illustration\label{NUM}}

To illustrate the numerical potential of our asymptotic formulas in
practice, we finish this article with a brief section comparing exact
(Monte-Carlo-simulated) option prices and IVs with the asymptotics we have
derived. Formulas such as (\ref{Eye}) can be used to calibrate various
parameters which might be linked explicitly or empirically to $\lambda
_{1}\left( 1\right) $, assuming $H$ is known. We refer to the numerics in
our prior work in \cite{GVZ} for details on what can be done, leaving to the
interested reader any details of how to translate the ideas therein which
are for extreme strike asymptotics to the small time case. \cite{GVZ} also
contains a description of how to simulate the fBm-driven models of interest
to us, for Monte-Carlo purposes, as alluded to in Remark \ref{R:KL}; we do
not repeat this information here.

Our results in the at-the-money case are presumably harder to exploit along
these lines because they depend on moment statistics $\mu _{1/2}$ and $\mu
_{3/2}$ (Remark \ref{R:intv}), which are not explicitly related to model
parameters. An exception to this observation is in the case of models with a
volatility scale parameter $\sigma $, by which we mean that one replaces
model (\ref{E:svm}) with 
\begin{equation}
dS_{t}=rS_{t}dt+\sigma \left\vert X_{t}\right\vert S_{t}dW_{t}.
\label{withsigma}
\end{equation}%
Here the parameter $\sigma $ is rather inoccuous since, by self-similarity
of $\left\vert X\right\vert $, this $\sigma $ can be absorbed as a linear
time change, but it represents a convenient parameter for tuning a model to
realistic time-scales and volatility levels. We will use this device in this
section. In particular, at the money, it is easy to see from Theorem \ref%
{T:aiv} that one has%
\begin{equation*}
I(T,s_{0})=\sigma ~\mu _{1/2}~T^{H}+\frac{\sigma ^{3}}{24}T^{3H+1}\left[
\left( \mu _{1/2}\right) ^{3}-\mu _{3/2}\right] +O\left( T^{5H+2}\right)
\end{equation*}%
where $\mu _{1/2}$ and $\mu _{3/2}$ are given in Remark \ref{R:intv}. Thus
at-the-money IV asymptotics can be used to calibrate $\sigma $ in model (\ref%
{withsigma}). We do not comment on this further herein.

Instead, we provide a numerical analysis of our results' use in $H$'s
calibration. Indeed, the reference \cite{GVZ} contains an effort to
calibrate $H$ itself, when other parameters have been estimated by other
means, but left some stones unturned. We found therein that $H$ calibration
can be relatively successful in some cases in practice, though this is not
necessarily backed up by any asymptotic theory. In this section we show
instead how model-free results such as Corollary \ref{C:fiv} and Theorem \ref%
{T:recover} provide excellent calibration of $H$ in many cases. We choose to
present this in the at-the-money case for two reasons. First, it illustrates
the model-free framework, since the results we obtain are not sensitive to
the values of $\mu _{1/2}$ and $\mu _{3/2}$. Second, in practice, liquidity
is low for options away from the money near maturity, which all but dictates
the use of at-the-money IV.

The setup we use is that of model (\ref{withsigma}) with $X=$ fBm, $r=0$,
and $\sigma =3$. The choice of $\sigma $ is tailored to provide a realistic
volatility level after 1 or 2 weeks, with time measured in years.
Specifically, a practicioner may simply select the desired magnitude of $%
\sigma $ by matching it to the mean magnitude of volatility in (\ref%
{withsigma}) via the formula%
\begin{equation*}
\mathbf{E}\left[ \sigma \left\vert X_{t}\right\vert \right] =\sigma t^{H}%
\sqrt{2/\pi }.
\end{equation*}%
For example, with $H=0.6$ and $\sigma =3$ we get $\mathbf{E}\left[ \sigma
\left\vert X_{t}\right\vert \right] \approx 0.22$ after one week ($%
t=7/365\approx 0.019$,) and $\mathbf{E}\left[ \sigma \left\vert
X_{t}\right\vert \right] \approx 0.34$ after one week ($t=14/365\approx
0.038 $), which could represent a realistic scenario for a volatile
short-term bond market. Values of $\sigma $ closer to unity result in much
smaller volatility values near maturity; these allow for an extremely sharp
fit between theoretical call and IV values and our asymptotics, but would
typically be unrealistically small, hence our choice of $\sigma =3$.

Before using Theorem \ref{T:recover}, a first question might be whether it
would not be sufficient to use an asymptotic theory for call prices to
estimate parameters. The use of IV over option prices has been advocated in
many articles, including many of the ones cited herein, but the question is
still legitimate since one rarely sees evidence in the literature that this
is indeed preferable in practice. The two images in Figure 1 compare the fit
between our asymptotic formulas (Corollary \ref{C:bis} and Theorem \ref%
{T:aiv}) and exact (simulated) call and IV values for times from 1 day to 2
weeks.

\begin{figure}[h]
\includegraphics[scale=0.25]{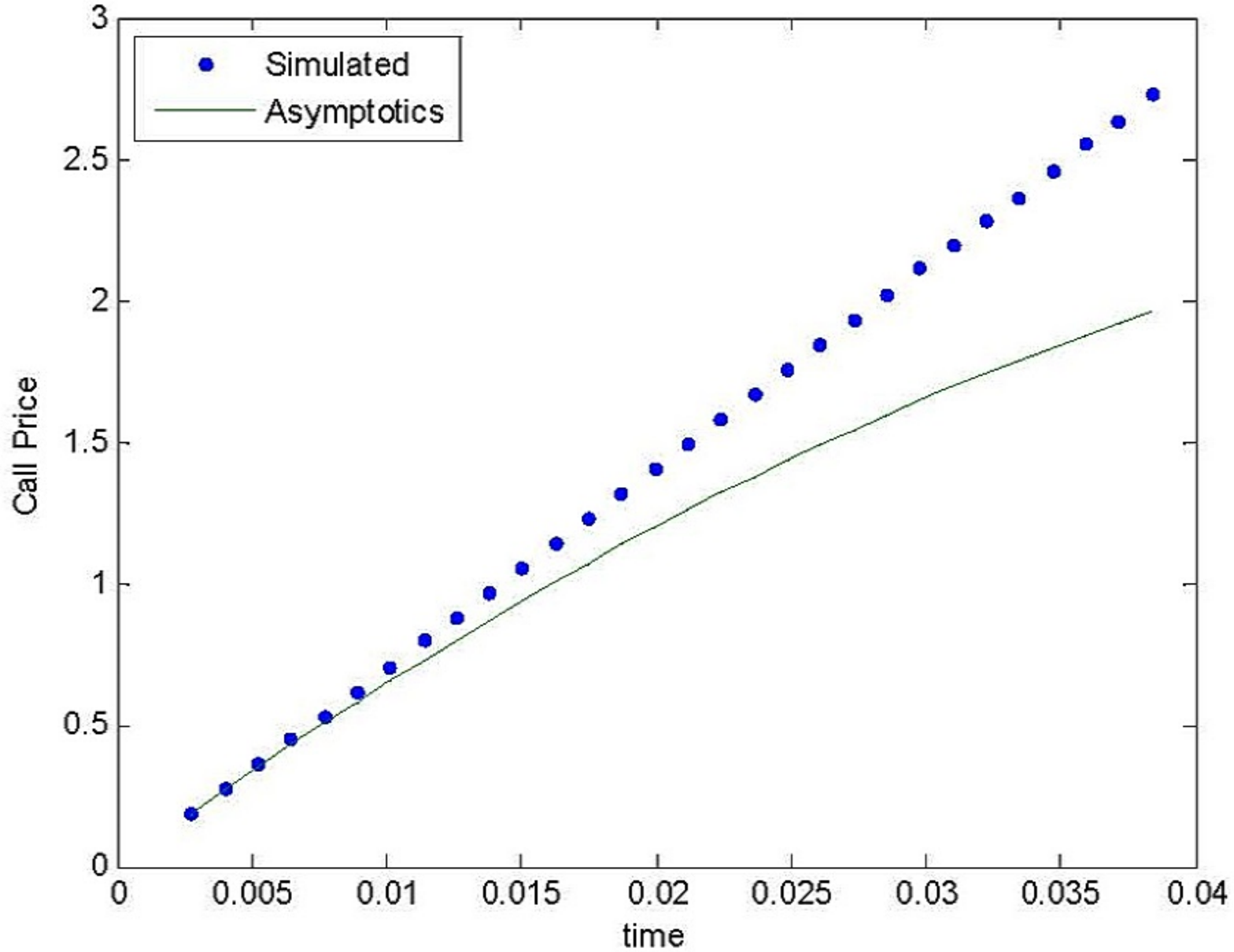}%
\includegraphics[scale=0.25]{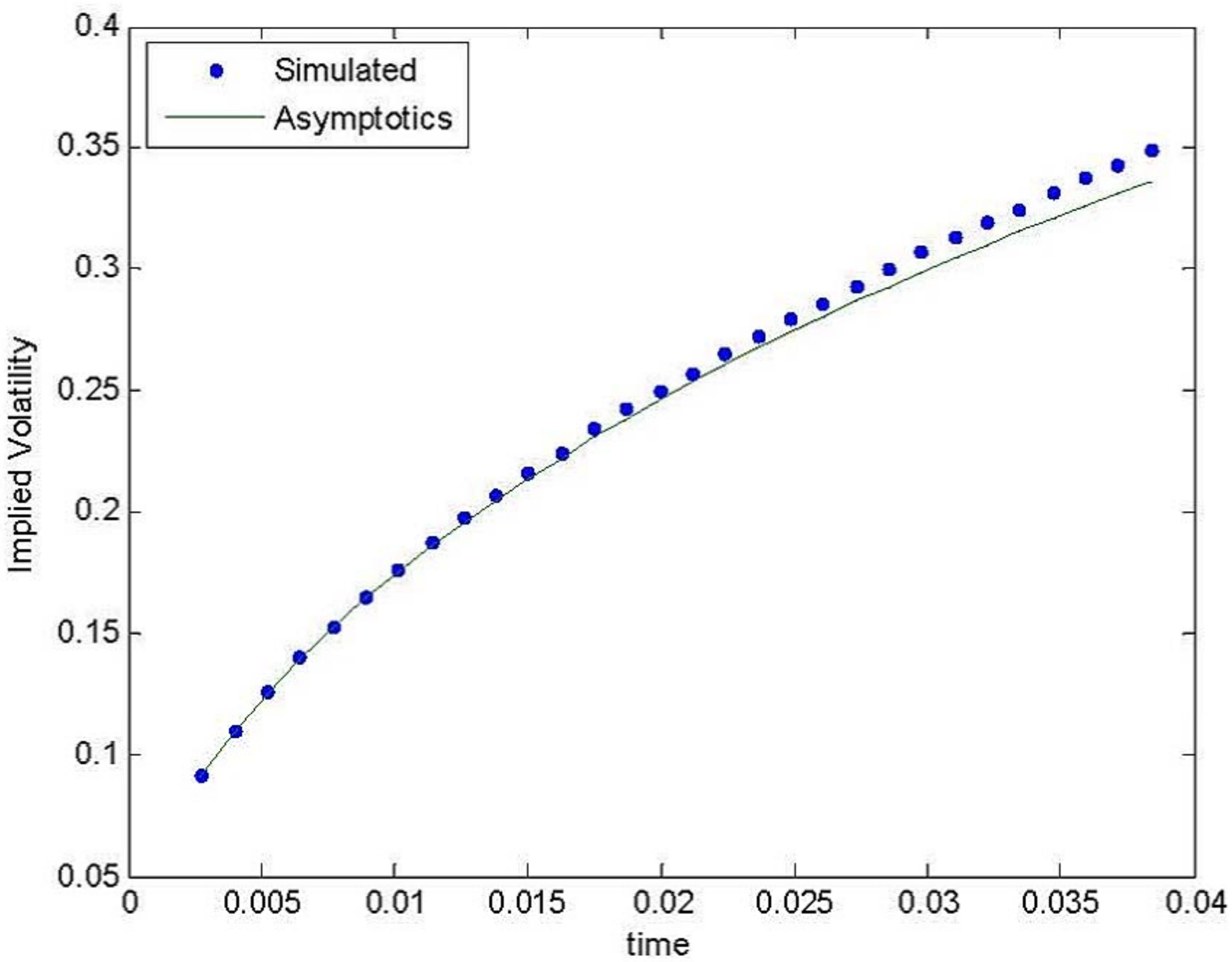}
\caption{Call (left) and IV (right) with $\protect%
\sigma =0.3$, $t\in \lbrack $1 day; 2 weeks$]$, $H=0.51$}
\label{fig1}
\end{figure}

We chose the extreme case $H=0.51$ because, as it turns out, the
asymptotics' accuracy increase as $H$ increases. We see from the above that
the IV asymptotics are accurate at a roughly $5\%$-error level for more than
10 days, and remains fairly accurate up to 2 weeks, while the call
asymptotics are only accurate at a $5\%$-error level for 2 days, and
deteriorate significantly thereafter. Other values of $H$ show similar
pictures. The choice to use IV over call prices for calibration purposes in
small time is clear. This can of course be verified rigorously on our
formulas since our coefficients can be computed numerically as well; this is
omitted from our study. The four pictures in Figure 2 show the extremely sharp fit
of IV asymptotics over two weeks as $H$ increases, as we mentioned.

\begin{figure}[h]
\includegraphics[scale=0.25]{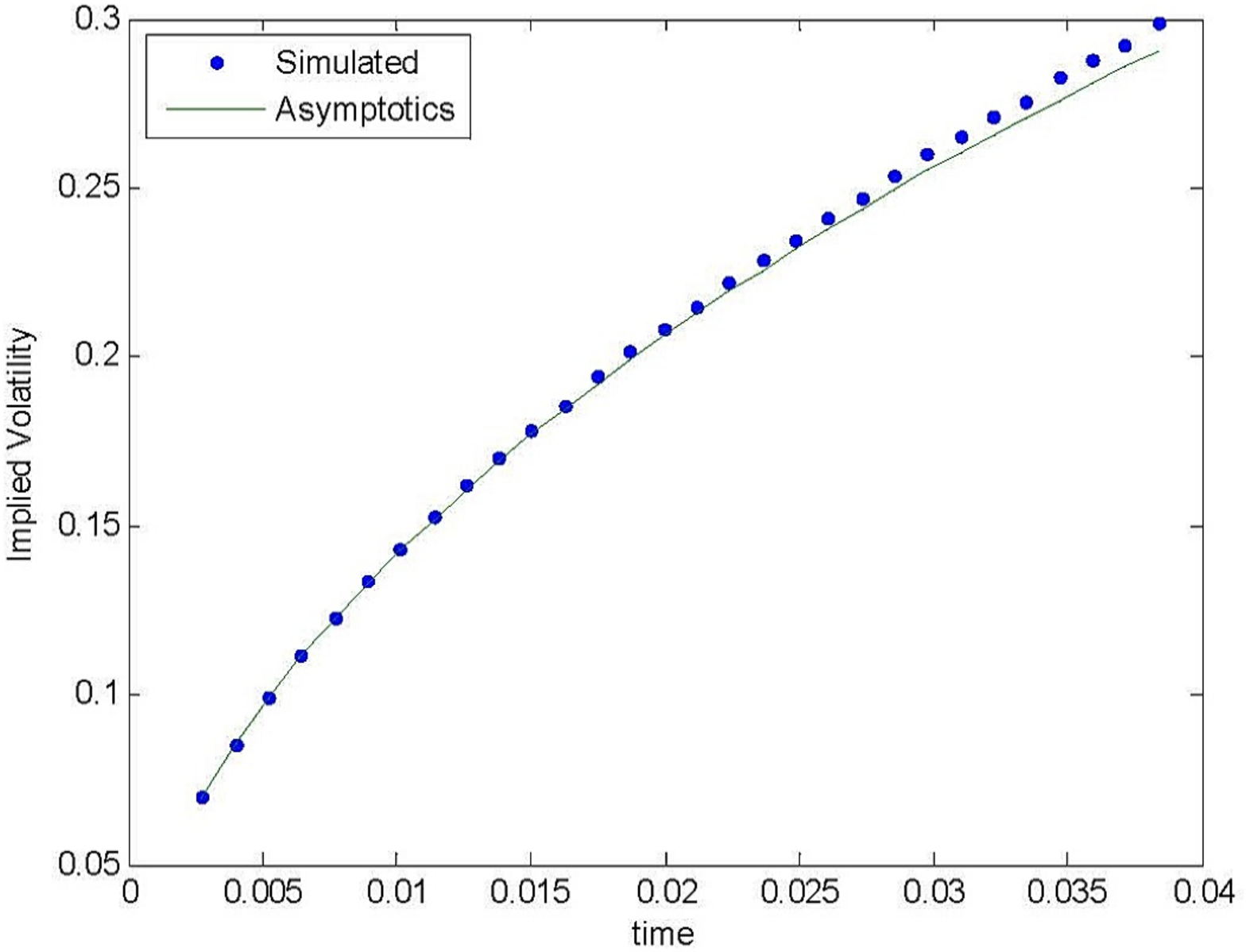}%
\includegraphics[scale=0.25]{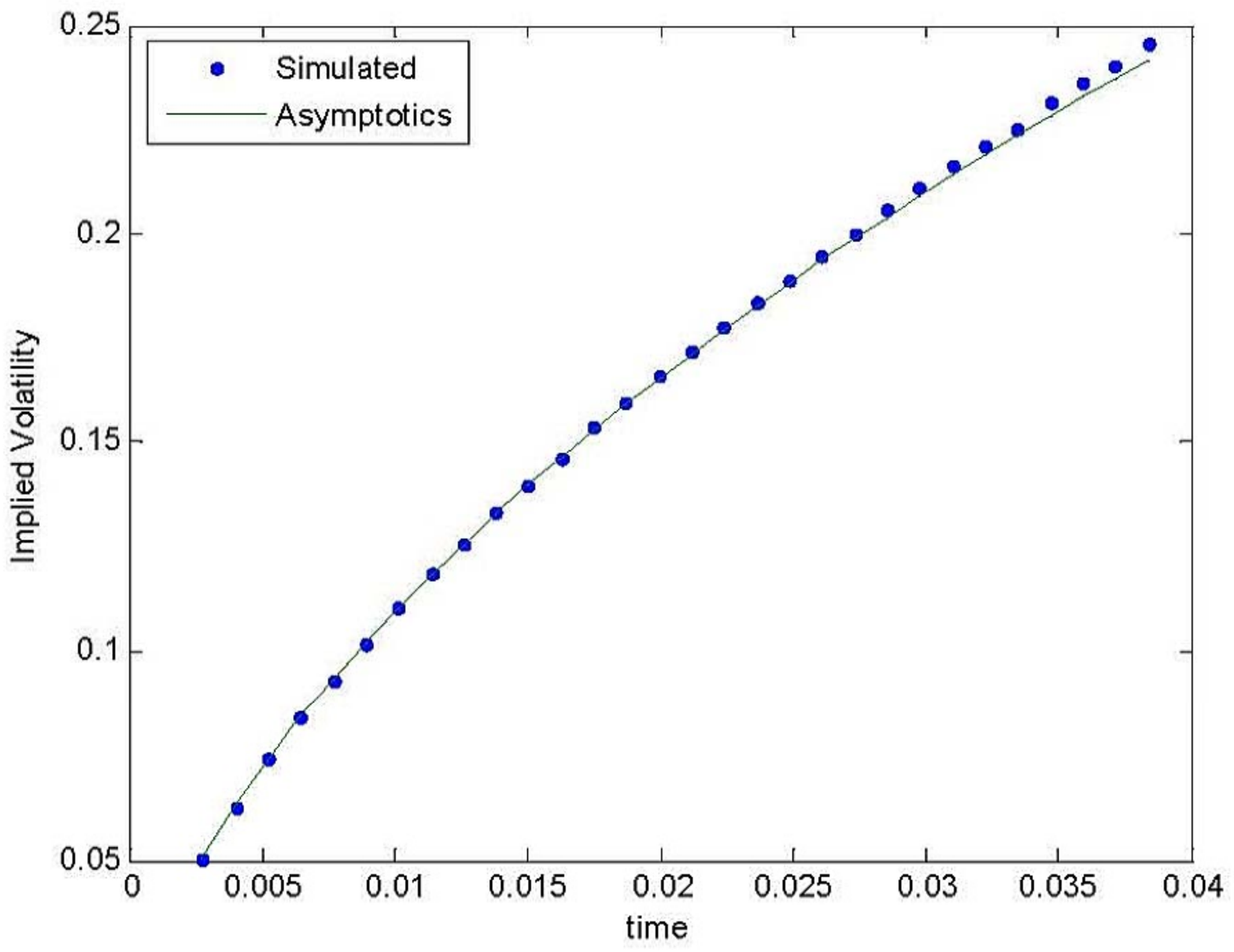}\newline
\includegraphics[scale=0.25]{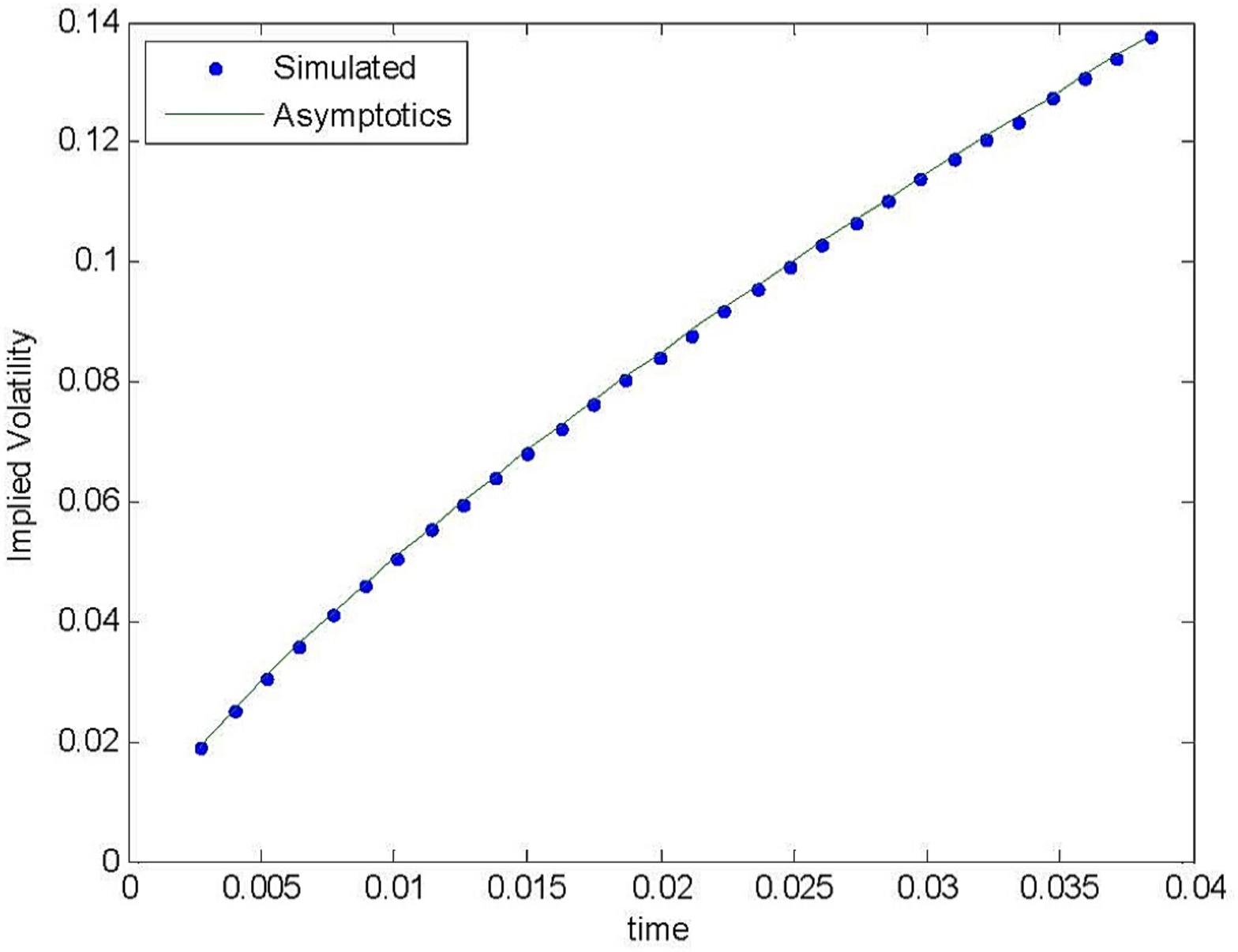}%
\includegraphics[scale=0.25]{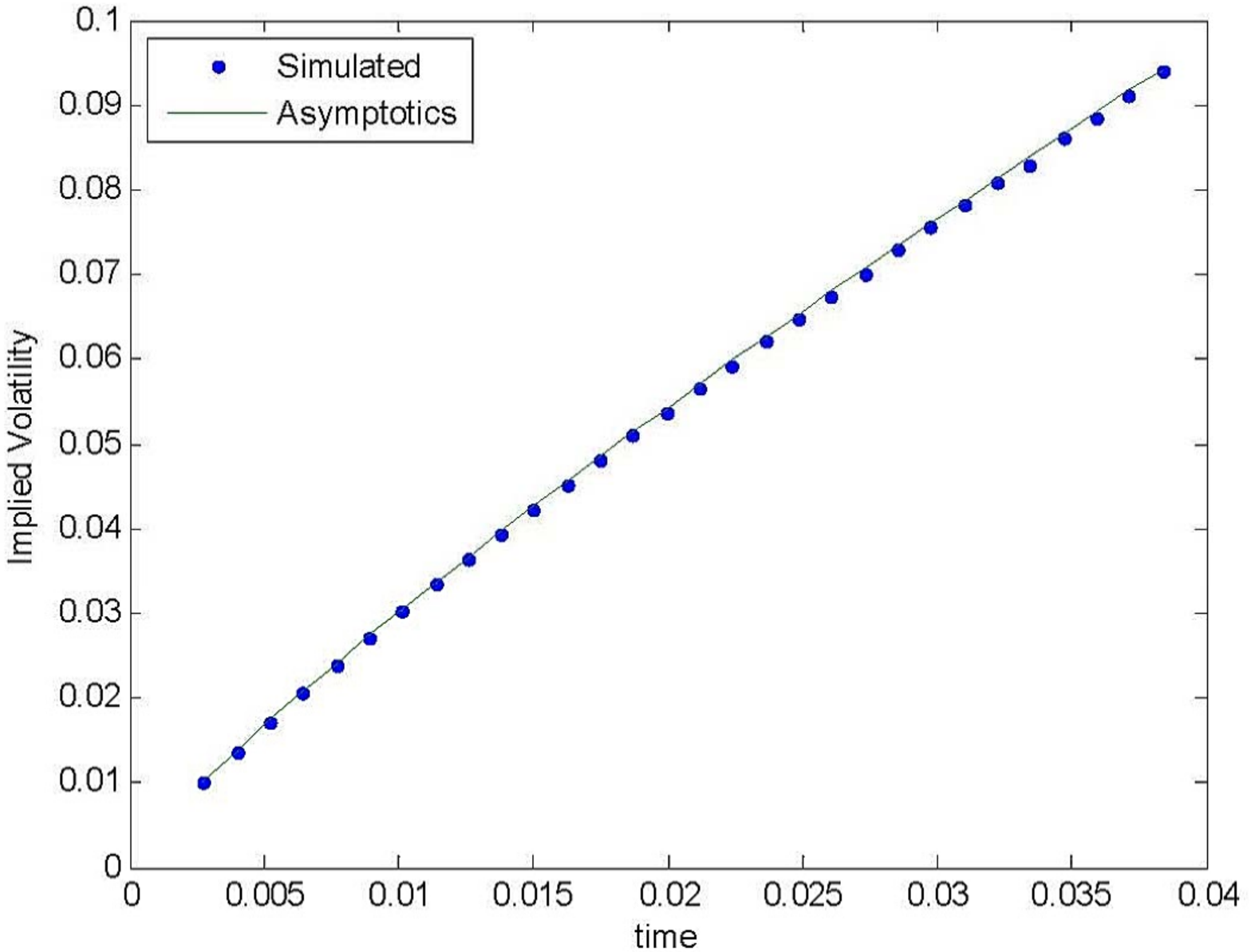}%
\caption{IV with $\protect%
\sigma =0.3$, $t\in [$1 day; 2 weeks$]$, $H=0.55,0.60$ (top left and right), $H=0.75,0.85$ (bottom left and right).}
\label{fig2}
\end{figure}

Since liquidity decreases as time to maturity decreases, it is desirable to
use the largest possible time $t_{0}$ such that the relative error in IV
approximation does not exceed a given error level, say $1\%$ which would be
a high level of accuracy. The table below give an idea of what this means in
practice, by computing $t_{0}$ for a $1\%$ level in the above realistic
cases: with 
\begin{equation*}
t_{0}=\max \left\{ t:\frac{\text{simulated IV }\left( t\right) -\text{%
asymptotic IV }\left( t\right) }{\text{simulated IV }\left( t\right) }%
<0.01\right\}
\end{equation*}%
we find :%
\begin{equation*}
\begin{tabular}{|l|l|l|l|l|l|}
\hline
$H$ & $0.51$ & $0.55$ & $0.60$ & $0.75$ & $0.85$ \\ \hline
$t_{0}$ in days & $2.3$ & $4.6$ & $10.4$ & $14$ & $14$ \\ \hline
\end{tabular}%
.
\end{equation*}%
These values of $t_{0}$ could be considered as rather conservative, due to
the choice of $1\%$ accuracy; practitioners may decide to choose a slightly
more liberal level. This is evident from the last tables below, in which we
show the result of the calibration of $H$ from exact (simulated) option
prices, via Theorem \ref{T:recover}.%
\begin{eqnarray*}
T &=&\text{ 1 day} \\
&&%
\begin{tabular}{|l|l|l|l|l|l|l|}
\hline
$H$ used in simulation & $0.50$ & $0.51$ & $0.55$ & $0.60$ & $0.75$ & $0.85$
\\ \hline
$H$ calibrated from IV via Theorem \ref{T:recover} & $0.50$ & $0.51$ & $0.55$
& $0.60$ & $0.75$ & $0.85$ \\ \hline
\end{tabular}%
\end{eqnarray*}%
\begin{eqnarray*}
T &=&\text{ 2 days} \\
&&%
\begin{tabular}{|l|l|l|l|l|l|l|}
\hline
$H$ used in simulation & $0.50$ & $0.51$ & $0.55$ & $0.60$ & $0.75$ & $0.85$
\\ \hline
$H$ calibrated from IV via Theorem \ref{T:recover} & $0.50$ & $0.51$ & $0.55$
& $0.60$ & $0.75$ & $0.85$ \\ \hline
\end{tabular}%
\end{eqnarray*}%
\begin{eqnarray*}
T &=&\text{ 7 days} \\
&&%
\begin{tabular}{|l|l|l|l|l|l|l|}
\hline
$H$ used in simulation & $0.50$ & $0.51$ & $0.55$ & $0.60$ & $0.75$ & $0.85$
\\ \hline
$H$ calibrated from IV via Theorem \ref{T:recover} & $0.50$ & $0.50$ & $0.55$
& $0.60$ & $0.75$ & $0.85$ \\ \hline
\end{tabular}%
\end{eqnarray*}%
\begin{eqnarray*}
T &=&\text{ 14 days} \\
&&%
\begin{tabular}{|l|l|l|l|l|l|l|}
\hline
$H$ used in simulation & $0.50$ & $0.51$ & $0.55$ & $0.60$ & $0.75$ & $0.85$
\\ \hline
$H$ calibrated from IV via Theorem \ref{T:recover} & $0.50$ & $0.50$ & $0.54$
& $0.59$ & $0.75$ & $0.85$ \\ \hline
\end{tabular}%
\end{eqnarray*}%
In all cases, even with a 14-day time to maturity, the error in $H$%
-calibration is no greater than one hundredth (less than $2\%$ relative
error). The only difficulty we experience appears to be in differentiating
between a model with Brownian scaling ($H=0.50$, no memory in the
volatility) and a model with $H>0.50$, except for the very short times to
maturity $t=1,2$ days. If liquidity at those levels is adequate, as it may
be in heavily traded bond markets, then our calibration can be used with
such short horizons. Otherwise a maturity of one week is preferable,
particularly for self-similarity indices which are not too close to $0.50$.
A maturity of two weeks will work in all cases for scenarios where one is
satisfied with a possible error of one hundredth on $H$ calibration; this
could be a realistic accuracy level for many users of stochastic volatiltiy
models who are currently not using any self-similarity or long-memory
assumptions.

\end{document}